\documentclass[Conference]{IEEEtran}
\usepackage{tabularx}
\usepackage{makecell}

\usepackage{enumitem}
\usepackage{url}
\usepackage{graphicx}
\usepackage{subcaption}

\usepackage{amsfonts}
\usepackage{mathtools}
\usepackage{multicol}
\usepackage[utf8]{inputenc}
\usepackage{newunicodechar}
\newunicodechar{−}{\textminus}
\usepackage{booktabs}
\usepackage{textcomp}
\usepackage{comment}
\usepackage{siunitx} 

\usepackage{xcolor}
\usepackage{subcaption}

\usepackage{mathtools}

\usepackage{stfloats}
\usepackage[below]{placeins}
\usepackage{float}
\usepackage{cite}

\usepackage{pifont}

\usepackage{amssymb}

\usepackage{multirow}
\usepackage{amsmath}
\usepackage{csquotes}
\usepackage{color,soul}
\usepackage[bookmarks=false]{hyperref}
\usepackage{algorithm}
\usepackage{algorithmic}

\makeatletter

\ifCLASSINFOpdf
\else
\fi

\begin{document}

\title{

Energy Efficient Near-Threshold Voltage Massive MIMO Computing
}

\author{\IEEEauthorblockN{
MD Salman Sakib\IEEEauthorrefmark{1},
Mehdi Safarpour\IEEEauthorrefmark{2},
Pejman Ghasemzadeh\IEEEauthorrefmark{1},
Shahriar Shahabuddin\IEEEauthorrefmark{1}
\\}

\IEEEauthorblockA{\IEEEauthorrefmark{1}Oklahoma State University, Stillwater, OK, USA\\
\IEEEauthorrefmark{2}University of Oulu, Oulu, Finland\\
}

\IEEEauthorblockA{ Email: \IEEEauthorrefmark{1}md\_salman.sakib@okstate.edu,
\IEEEauthorrefmark{2}mehdi.safarpuor@oulu.fi,
\IEEEauthorrefmark{1}pejman.ghasemzadeh@okstate.edu,
\IEEEauthorrefmark{1}shahriar.shahabuddin@okstate.edu
}}

\maketitle

\begin{abstract}

 Near-Threshold Computing (NTC) offers a substantial energy savings, i.e., up to 10×, for digital circuits. However, due to Process, Voltage, and Temperature (PVT) variations, low voltage operating digital processors face increased susceptibility to timing errors. This work equips conventional massive MIMO algorithms with computational error detection mechanisms, enabling reliable operation down to near-threshold voltages. While conventional hardware based Timing Error Detection (TED) systems introduce substantial overhead, this work modifies MIMO algorithms to seamlessly integrate low overhead Algorithm Based Fault Tolerance (ABFT). A co-simulation framework combining transistor level SPICE simulations and workload aware fault injection is developed to investigate reliability, Bit Error Rate (BER), energy savings, and overheads of modified algorithms operating at low voltages. The transistor-level characterization was performed in PSpice using generic Level-1 Shichman–Hodges CMOS transistor models with Monte Carlo parameter variations. No specific CMOS fabrication process node was assumed. Consequently, the reported energy savings are analytical estimates and are not specific to a particular CMOS fabrication technology. We evaluate Gauss Seidel (GS), ADMM based infinity norm detection (ADMIN), and Conjugate Gradient (CG) detectors using circuit derived fault models. Results demonstrate remarkable algorithmic resilience. Although undetected computational errors in some trials near the Point of First Failure (PoFF) were observed, BER increases by only 0.7\% to 1.4\%, while beyond PoFF, ABFT reported nearly all erroneous cases. The proposed checksum based ABFT detects most arithmetic and memory errors, with an estimated overhead below 2\% for a 64~×~16 MIMO system, decreasing further for larger configurations. Analytical energy estimates indicate approximately 70\% dynamic energy savings relative to nominal voltage operation with tolerable overheads from the error detection mechanism. These results highlight the potential of combining inherent MIMO fault tolerance with lightweight ABFT for energy efficient and reliable computing at aggressively reduced voltages.

\end{abstract}

\begin{IEEEkeywords}
baseband processing, wireless SoC, fault injection, circuit reliability, checksum-based error detection, voltage scaling.
\end{IEEEkeywords}

\IEEEpeerreviewmaketitle

\vspace{-0.2cm}
\section{Introduction}

Massive multiple-input multiple-output (MIMO) improves wireless communication efficiency by using a large number of antennas. This allows the system to serve many users at the same time while making better use of the available spectrum~\cite{larsson2014massive, lu2014overview}. In an uplink Massive MIMO receiver, signals transmitted by multiple users must be separated at the base station. As the number of base-station antennas and active users
increases, the computational cost of symbol detection also increases
considerably~\cite{marzetta2015massive, ngo2013energy}. Consequently, the design of massive MIMO detectors involves a trade-off between performance and complexity. Hence, this requires consideration of increasingly important power and thermal reliability issues in constrained baseband processors~\cite{albreem2019massive, trotobas2020review, albreem2021low}. A wide range of detection algorithms has been developed to reduce the
computational burden of massive MIMO detection~\cite{yang2015fifty}. Iterative and approximate approaches such as Gauss--Seidel (GS) and Conjugate Gradient (CG) avoid the high computational cost of direct matrix inversion~\cite{albreem2019massive, shahabuddin2025massive}. Other approaches, including
ADMM-based infinity-norm detection (ADMIN) is another approach that incorporates additional constellation constraints to improve detection performance. It maintains an iterative computational structure while enforcing these constraints~\cite{sakib2025vlsi, shahabuddin2021admm}. Although these detectors reduce algorithmic complexity, they still involve intensive arithmetic and memory operations~\cite{sakib2026massive}. This results in considerable energy consumption during receiver processing.

Due to quadratic relation of digital dynamic power consumption with supply voltage, $P\sim V^2$, scaling supply-voltage is considered an attractive approach for reducing the energy consumption of digital circuits~\cite{pillai2001real,bellal2024gas}. Even more interesting, lowering voltage to the near-threshold voltage of transistors promises to substantially reduce the energy consumption of digital hardware, ideally up to 10×, with silicon proofs of 2× to 5× \cite{aliagha2025scissors,dreslinski2010near}.
 However, as the supply voltage approaches the transistor
threshold voltage, reduced timing and logic margins increase sensitivity to Process, Voltage, and Temperature (PVT) variations~\cite{beiu2003vlsi, burnett1995statistical, shahabuddin2025self}. Those unfortunately can not be modeled using conventional static timing analysis approach for low voltage regimes. These computational errors can corrupt both arithmetic outputs and stored intermediate values. For the case of MIMO, those errors propagate across multiple iterations of the detector and may manifest as amplified noise in output. Thus, aggressive voltage scaling introduces a
fundamental trade-off between energy efficiency and computational
reliability~\cite{chippa2010scalable}. Moreover, uniform
random fault models do not capture the non-uniform voltage sensitivity of different circuit blocks and bit positions~\cite{hamdioui2004testing}.

\subsection{Related Work}
\label{subsec:related_work}

\begin{figure}[t]
    \centering
    \includegraphics[width=0.99\columnwidth]{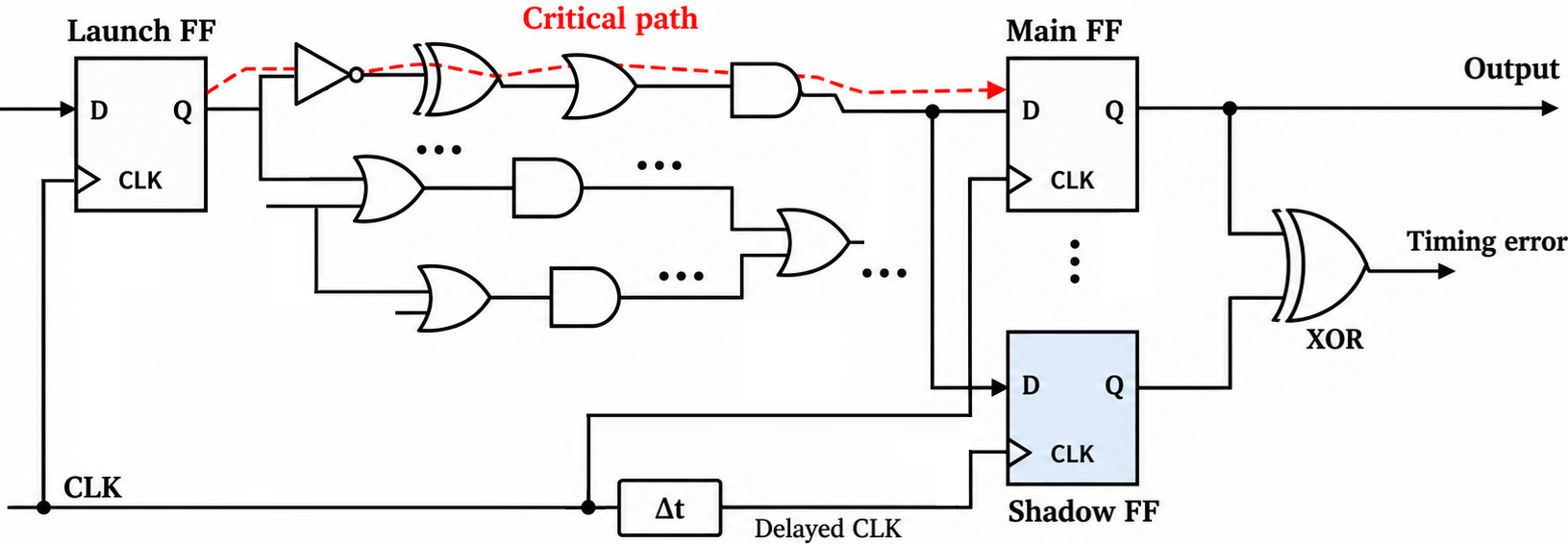}
    \caption{A simplified Timing Error Detection circuitry that is added to error susceptible delay path(s) for detecting timing errors induced by low voltage operating points \cite{safarpour2021algorithm}.}
    \label{fig:TED}
\end{figure}

Prior massive MIMO research has extensively investigated iterative and approximate-inversion detectors to reduce computational complexity ~\cite{albreem2019massive}. Although several hardware-oriented implementations have been reported, but reduced voltage operation and associated reliability of arithmetic operations under aggressive voltage scaling has received comparatively almost no attention.

For low voltage operating digital processors, including matrix engines, at the circuit level,timing-error monitors known as Timing Error Detection (TED) circuits,
critical-path replicas, calibration, and redundancy techniques have been investigated~\cite{zhang2018thundervolt,dreslinski2010near,aliagha2025scissors}. An error detection element of TED system is depicted in Fig.~\ref{fig:TED} where a shadow flip-flop captures late arrival signal and comparison with original sampled signal reveals existence of an timing error. However, incorporating fine-grained TED into matrix-processing engines requires additional sampling registers, comparison logic, and control circuitry across numerous multiply–accumulate (MAC) datapaths. These additional hardware components increase area, power consumption, and design complexity. Prior Razor-based systolic-array implementations have reported approximately 30\% area overhead. Protecting only selected critical paths can reduce this overhead, but may leave some timing errors undetected.\cite{ernst2003razor}. On the other hand, Algorithm-Based Fault Tolerance (ABFT) offers an algorithm-level approach for detecting computational errors. ABFT exploits mathematical relationships within matrix and vector operations. These properties make ABFT well suited for massive MIMO processing~\cite{ rinkinen2025near,safarpour2021high}. 

Recent work on near-threshold massive MIMO computing integrated ABFT with
a Newton-iteration detector on an undervolted FPGA accelerator~\cite{rinkinen2025near}. The
approach demonstrated low-overhead computational error detection and
significant power savings without detailed cell-level characterization.
However, that work primarily focused on datapath-level errors and did not explicitly model memory reliability or nonuniform bit-level circuit vulnerability in approach similiar to \cite{xie2025realm}. In contrast, we considered a cross-layer approach in which transistor-level characterizations of datapath, i.e., MAC units, and the on-chip memory, i.e, SRAM cells, are converted into voltage-dependent 16-bit vulnerability profiles for system-level fault injection. We then use computational profiling to relate the circuit-level vulnerability characteristics to the arithmetic workloads of GS, ADMIN, and CG. Finally, we incorporate checksum-based ABFT in which checksum information is propagated through the detector and verified once at the final recovered output. This framework therefore connects circuit-level reliability characteristics, detector-specific fault exposure, and final-output error detection within a unified massive MIMO reliability analysis.

The main contributions of this work are summarized as follows:

\begin{itemize}

    \item We perform transistor-level reliability characterization of the
 memory and datapath blocks used in massive MIMO detection under
    voltage scaling and device variations. We evaluate an 8-bit ripple-carry adder, an 8-bit Wallace-tree multiplier, and a 6T SRAM cell in PSpice. We then use the measured circuit behavior to construct voltage-dependent 16-bit vulnerability profiles for system-level reliability analysis.

    \item We develop a workload-guided fault-exposure methodology for the
    GS, ADMIN, and CG detectors. We use dynamic operation counting to quantify the arithmetic workloads of the three detectors. Our analysis identifies multiplication and addition as the dominant computational operations.
    We then use the detector-specific operation counts to account for the
    different cumulative fault exposures of GS, ADMIN, and CG.

    \item We develop a system-level fault-injection framework that maps the
    PSpice-derived reliability characteristics to the corresponding
    arithmetic and memory operations of the massive MIMO detectors. During each arithmetic fault event, we select an affected result word and bit position based on the corresponding circuit-derived vulnerability profile. We then flip the selected bit in the signed 16-bit fixed-point representation. Because fault events can occur repeatedly during detector
    execution, multiple bit corruptions may accumulate within a MIMO trial.
    We evaluate SRAM faults separately using the measured voltage-dependent
    SRAM failure profile.

    \item We incorporate a detection-only checksum-based ABFT mechanism in
    which checksum information is appended to the detector state and
    propagated through the GS, ADMIN, and CG computations. We perform a
    single checksum verification at the final recovered detector output and
    classify corrupted outputs as detected or silent depending on the final
    checksum comparison.

    \item We evaluate the proposed cross-layer framework using final-output
    error statistics and Bit Error Rate (BER) simulations for GS, ADMIN, and CG across the
    considered supply-voltage range. We identify the arithmetic point of
    first failure, examine BER robustness near this operating point, and
    compare detector-dependent error-propagation behavior. We also evaluate
    SRAM-originated errors separately and analyze the computational and
    estimated energy overhead of the checksum-based ABFT mechanism as the
    MIMO system size increases.

\end{itemize}

\section{System Model}
\label{sec:system_model}

\begin{figure}[t]
    \centering
    \includegraphics[width=0.92\columnwidth]{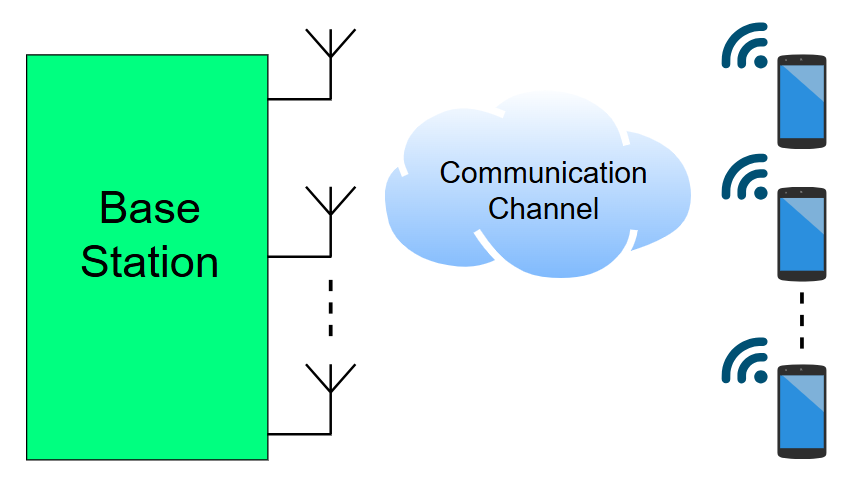}
    \caption{Uplink Massive MIMO system with $U$ users and a $B$-antenna base station.}
    \label{fig:massive_mimo_system}
\end{figure}

The uplink of a massive multiuser MIMO system is considered, where $U$ single-antenna users are simultaneously served by a base station (BS) equipped with $B$ antennas, with $B \geq U$. Each user maps its information
bits to a complex-valued modulation symbol drawn from a constellation
set $\mathcal{O}$. The symbols transmitted by all users are collected
in the vector $\mathbf{x}\in\mathbb{C}^{U}$, while the signal received
at the BS is represented by $\mathbf{y}\in\mathbb{C}^{B}$. Assuming a
frequency-flat channel, the uplink baseband model is

\begin{equation}
    \mathbf{y}=\mathbf{H}\mathbf{x}+\mathbf{n},
    \label{eq:system_model}
\end{equation}
where $\mathbf{H}\in\mathbb{C}^{B\times U}$ denotes the channel matrix
and $\mathbf{n}\in\mathbb{C}^{B}$ is the additive white Gaussian noise
(AWGN) vector. Perfect channel state information (CSI) and synchronization
are assumed at the BS. The receiver uses the observed vector $\mathbf{y}$
and channel matrix $\mathbf{H}$ to estimate the transmitted symbol vector
$\mathbf{x}$~\cite{albreem2019massive}. Fig.~\ref{fig:massive_mimo_system} illustrates the considered uplink
Massive MIMO architecture. A set of $U$ single-antenna users transmit
simultaneously over the wireless channel to a BS equipped with $B$
antennas. The BS uses the received vector together with the estimated
channel matrix to recover the transmitted symbols using the considered
minimum mean-square-error (MMSE), GS, ADMIN, and CG detectors. For the considered detectors, it is convenient to define the
regularized Gram matrix and matched-filter vector as

\begin{equation}
    \mathbf{A}
    =
    \mathbf{H}^{H}\mathbf{H}
    +
    \frac{N_0}{E_s}\mathbf{I},
    \qquad
    \mathbf{b}
    =
    \mathbf{H}^{H}\mathbf{y},
    \label{eq:gram_matched}
\end{equation}
where $N_0$ denotes the noise variance and $E_s$ is the average
transmitted symbol energy. These quantities form the basis of the
MMSE solution and are also used by the iterative GS and CG detectors.
After detection, the estimated entries are mapped to the nearest
constellation points to obtain the final detected symbol vector.

\subsection{MMSE Detection}

The MMSE detector provides a linear estimate
while accounting for both multiuser interference and receiver noise.
Unlike zero-forcing detection, the regularization term prevents excessive
noise enhancement when the channel matrix is poorly conditioned
~\cite{wubben2004near}. Using the definitions in
(\ref{eq:gram_matched}), the MMSE estimate is

\begin{equation}
    \hat{\mathbf{x}}_{\mathrm{MMSE}}
    =
    \mathbf{A}^{-1}\mathbf{b}
    =
    \left(
    \mathbf{H}^{H}\mathbf{H}
    +
    \frac{N_0}{E_s}\mathbf{I}
    \right)^{-1}
    \mathbf{H}^{H}\mathbf{y}.
    \label{eq:mmse_detector}
\end{equation}
The resulting vector is subsequently quantized to the nearest point in
$\mathcal{O}$. MMSE provides a useful performance reference for the
iterative detectors considered in this work. Its direct implementation,
however, requires inversion of a $U\times U$ matrix, motivating
lower-complexity iterative approaches for large massive MIMO systems.

\subsection{Gauss--Seidel Detection}

The GS detector avoids directly computing the complete
matrix inverse by iteratively solving the MMSE linear system
$\mathbf{A}\hat{\mathbf{x}}=\mathbf{b}$. The matrix $\mathbf{A}$ is
decomposed as $\mathbf{A}=\mathbf{D}+\mathbf{L}+\mathbf{U}$, where
$\mathbf{D}$ contains the diagonal entries and $\mathbf{L}$ and
$\mathbf{U}$ are the strictly lower- and upper-triangular components,
respectively~\cite{ahmad2022high}. The GS update is

\begin{equation}
    \hat{\mathbf{x}}^{(k+1)}
    =
    (\mathbf{D}+\mathbf{L})^{-1}
    \left(
    \mathbf{b}
    -
    \mathbf{U}\hat{\mathbf{x}}^{(k)}
    \right).
    \label{eq:gs_detector}
\end{equation}
Each iteration reuses the most recently available symbol estimates,
allowing the solution to progressively approach the MMSE result.
Consequently, the number of iterations provides a direct trade-off
between detection accuracy and computational workload. The repeated
matrix--vector multiplications and additions also make GS relevant to
the computational fault-exposure analysis considered later in this work.

\subsection{ADMM-Based Infinity-Norm Detection}

The ADMIN detector accounts for the finite modulation alphabet through a convex relaxation~\cite{sakib2025vlsi}. Specifically, it replaces the discrete constellation constraint with a convex box that encloses the constellation. In contrast to unconstrained linear
MMSE detection, ADMIN therefore solves a relaxed constrained detection
problem. Introducing an auxiliary variable $\mathbf{z}$, the formulation
can be written as 

\begin{equation}
\begin{aligned}
    \underset{\mathbf{x},\mathbf{z}}{\operatorname{minimize}}
    \quad &
    \frac{1}{2}
    \left\|
    \mathbf{y}-\mathbf{H}\mathbf{x}
    \right\|_2^2
    +g(\mathbf{z}),
    \\
    \text{subject to}\quad &
    \mathbf{x}=\mathbf{z},
\end{aligned}
\label{eq:admin_problem}
\end{equation}
where $g(\mathbf{z})$ represents the indicator function of the
constellation-enclosing box. A scaled ADMIN iteration is given by

\begin{equation}
\begin{aligned}
\mathbf{x}^{(k+1)}
&=
(\mathbf{H}^{H}\mathbf{H}+\beta\mathbf{I})^{-1}
\left[
\mathbf{H}^{H}\mathbf{y}
+\beta(\mathbf{z}^{(k)}-\boldsymbol{\lambda}^{(k)})
\right],\\
\mathbf{z}^{(k+1)}
&=
\Pi_{\mathcal{C}_{\mathcal{O}}}
\left(
\mathbf{x}^{(k+1)}+\boldsymbol{\lambda}^{(k)}
\right),\\
\boldsymbol{\lambda}^{(k+1)}
&=
\boldsymbol{\lambda}^{(k)}
-\gamma
\left(
\mathbf{z}^{(k+1)}-\mathbf{x}^{(k+1)}
\right),
\end{aligned}
\label{eq:admin_updates}
\end{equation}
where $\beta>0$ is the penalty parameter, $\gamma$ is the relaxation
parameter, and $\Pi_{\mathcal{C}_{\mathcal{O}}}(\cdot)$ denotes
projection onto the constellation box. With zero initialization, the
first ADMIN iteration corresponds to an MMSE-type update, while
subsequent iterations progressively enforce the box constraint.

\subsection{Conjugate-Gradient Detection}

The CG detector is another iterative approach for
solving the regularized linear system
$\mathbf{A}\hat{\mathbf{x}}=\mathbf{b}$ without explicitly forming
$\mathbf{A}^{-1}$. CG searches for the solution along mutually conjugate
directions and successively refines the estimated symbol vector
~\cite{wei2020learned}. Its primary estimate update is

\begin{equation}
    \hat{\mathbf{x}}^{(k+1)}
    =
    \hat{\mathbf{x}}^{(k)}
    +
    \alpha_k\mathbf{p}_k,
    \qquad
    \mathbf{p}_k^{H}\mathbf{A}\mathbf{p}_j=0,
    \quad k\neq j,
    \label{eq:cg_detector}
\end{equation}
where $\mathbf{p}_k$ denotes the conjugate search direction and
$\alpha_k$ controls the step along that direction. The residual and
search direction are updated at each iteration until the desired
iteration count or convergence condition is reached. Because CG relies
primarily on matrix--vector products rather than direct inversion, it is
well suited to large-dimensional massive MIMO detection. Its iterative structure creates a computational dataflow that differs from GS and ADMIN. This difference motivates its inclusion in the comparative evaluation of complexity and fault tolerance.

The three evaluated detectors provide different computational structures
for analyzing near-threshold hardware reliability. GS and CG iteratively
solve the regularized linear detection problem using different update
mechanisms, whereas ADMIN introduces an additional
constellation-aware box constraint through an ADMM-based formulation.
In the following sections, the arithmetic and memory operations associated
with these detectors are mapped to the circuit-derived vulnerability
characteristics of the multiplier, adder, and SRAM. This enables the
impact of near-threshold hardware faults to be evaluated at the
end-to-end massive MIMO detection level.

\section{Complexity Analysis Toolbox}

To identify the computational blocks that are most strongly exposed to
hardware faults during massive MIMO detection, an operation-counting
analysis was performed for the GS, ADMIN, and CG detectors. In this work,
the operation count is not used only as a conventional measure of
algorithmic complexity. Instead, it is used to determine how frequently
different arithmetic resources are exercised during detector execution.
A computational block that is activated more frequently encounters a
larger number of opportunities for circuit-level faults to affect the
algorithmic datapath. The resulting complexity profiles therefore provide
a basis for identifying the arithmetic blocks that require particular
attention in the subsequent reliability analysis. The analysis was performed for the considered $64\times16$ massive MIMO
configuration with 64-QAM modulation using the corresponding detector
implementations. The operation-counting toolbox dynamically traces the
executed numerical operations and records complex multiplications,
complex additions, real value multiplications, real value additions, divisions, and
square-root operations. Unlike a purely static source-code count, the
tracing procedure follows the actual executed computational paths and
loop iterations of each profiled detector.

Table~\ref{tab:detector_complexity} summarizes the measured arithmetic
and dataflow characteristics of the three detectors. Complex
multiplication and addition constitute the dominant arithmetic workload
in all cases. GS requires 21,512 complex multiplications and 21,255
complex additions, while ADMIN requires 18,808 and 19,520, respectively.
CG requires 18,320 complex multiplications and 19,053 complex additions.
The corresponding real-equivalent workloads are 89,360 multiplications
and 86,784 additions for GS, 78,800 and 77,960 for ADMIN, and 76,745
and 76,065 for CG. The equivalent counts are obtained by representing one complex
multiplication using four real multiplications and two real additions.
Accordingly, the real-equivalent multiplication count is

\begin{equation}
N_{\mathrm{mult}}^{\mathrm{eq}}
=
N_{\mathrm{RM}}+4N_{\mathrm{CM}},
\label{eq:eq_mult}
\end{equation}
where $N_{\mathrm{RM}}$ and $N_{\mathrm{CM}}$ denote the numbers of real
and complex multiplications, respectively. Similarly, one complex
addition corresponds to two real additions, giving

\begin{equation}
N_{\mathrm{add}}^{\mathrm{eq}}
=
N_{\mathrm{RA}}
+2N_{\mathrm{CA}}
+2N_{\mathrm{CM}},
\label{eq:eq_add}
\end{equation}
where $N_{\mathrm{RA}}$ and $N_{\mathrm{CA}}$ denote the numbers of real
and complex additions, respectively. Using Eqs.~\ref{eq:eq_mult} and~\ref{eq:eq_add}, complex multiplication
accounts for approximately $96.3\%$, $95.5\%$, and $95.5\%$ of the
real-equivalent multiplication workload of GS, ADMIN, and CG,
respectively. Similarly, complex arithmetic contributes approximately
$98.6\%$, $98.3\%$, and $98.3\%$ of their real-equivalent addition
workloads. These results show that the multiplier and adder datapaths are used repeatedly across all three detector architectures. As a result, they are especially important from a hardware-reliability perspective.

\begin{table}[t]
\centering
\caption{Computational complexity of the profiled Massive MIMO detectors.}
\label{tab:detector_complexity}
\footnotesize

\begin{tabular*}{\columnwidth}{@{\extracolsep{\fill}} l r r r}
\hline
\textbf{Metric} & \textbf{GS} & \textbf{ADMIN} & \textbf{CG} \\
\hline
Complex multiplications         & 21,512 & 18,808 & 18,320 \\
Complex additions               & 21,255 & 19,520 & 19,053 \\
Real multiplications            & 3,312  & 3,568  & 3,465  \\
Real additions                  & 1,250  & 1,304  & 1,319  \\
Divisions                       & 81     & 113    & 7      \\
Square-root operations          & 1,024  & 1,040  & 1,033  \\
\hline
Real-equivalent multiplications & 89,360 & 78,800 & 76,745 \\
Real-equivalent additions       & 86,784 & 77,960 & 76,065 \\
\hline
Dataflow operation nodes        & 10     & 11     & 20     \\
Maximum parallel width          & 3      & 3      & 3      \\
Critical-path stages            & 16     & 17     & 33     \\
Measured software time (s)      & 0.058482 & 0.054629 & 0.005685 \\
\hline
\end{tabular*}
\end{table}

A high operation count does not imply that an individual multiplication
or addition has a higher intrinsic probability of failure. Rather, a
larger number of executions increases the cumulative exposure of the
corresponding hardware resource to circuit-level faults. Each invocation
of an adder or multiplier represents another opportunity for a
voltage-dependent error to enter the computational datapath. This distinction becomes especially important during near-threshold-voltage operation. At these lower operating margins, circuits become more sensitive to device-level variations. Fault exposure is also influenced by the computational role of an
intermediate result. Outputs generated by arithmetic operations are
frequently consumed by subsequent matrix--vector operations and
iterative updates. A corrupted value can therefore propagate through
dependent computations and eventually influence the estimated symbol
vector. As a result, the system-level impact of an arithmetic fault depends on both operation frequency and computational dependency. Operation count alone does not fully determine fault significance. The comparative results also reveal differences in the computational
structures of the three detectors. GS exhibits the largest
real-equivalent arithmetic workload among the evaluated implementations,
followed by ADMIN and CG. However, the dataflow characteristics show a
different trend. GS and ADMIN contain 10 and 11 traced operation nodes
with critical-path depths of 16 and 17 stages, respectively, whereas CG
contains 20 operation nodes and a critical-path depth of 33 stages. All
three detectors exhibit a maximum parallel width of three. These
dataflow quantities represent the dependency structure observed by the
operation-counting toolbox and should not be interpreted as physical
hardware pipeline stages. Nevertheless, they provide additional
information about how intermediate results are connected and how an
erroneous value may propagate through subsequent computations.\newline
The differences among the three detectors therefore illustrate that
fault exposure cannot be characterized solely by a single complexity
metric. GS carries out the highest number of arithmetic operations, resulting in the greatest overall arithmetic exposure among the evaluated implementations. In contrast, CG shows a deeper computational dependency structure.
ADMIN lies between these cases in terms of both arithmetic workload and
dataflow depth. Although the three detectors differ in their computational structures, multiplication and addition are the most frequently used arithmetic operations in each. Therefore, the circuit-level reliability analysis focuses on the corresponding multiplier and adder blocks. The measured execution times of approximately $58.5$~ms,
$54.6$~ms, and $5.69$~ms for GS, ADMIN, and CG, respectively, are
reported only as references for the profiled software implementations.
These results are not considered indicators of hardware latency. Software execution time depends on the implementation environment and does not directly reflect hardware timing. For the present reliability analysis, the
operation counts and computational dependencies are therefore more
relevant than the measured software runtime. Based on the comparative complexity profiles, the circuit-derived
vulnerability characteristics of the multiplier and adder are mapped to
the corresponding arithmetic operations in the system-level reliability
model. This establishes a direct connection between transistor-level
fault behavior and the computational activity of the GS, ADMIN, and CG
detectors.The resulting workload guided analysis provides the basis for the circuit aware fault injection framework and checksum based ABFT error detection mechanism introduced in the following sections.

\section{Design and Simulation of Computational and Memory Circuits}
\label{sec:computational_memory_blocks}

To obtain circuit-level reliability characteristics for near-threshold-voltage
operation, the computational and memory components considered in this work were
implemented and simulated in PSPICE. The evaluated blocks include an adder, a
multiplier, and a 6T SRAM bit-cell. The PSpice characterization uses generic Level-1 Shichman–Hodges NMOS and PMOS transistor models with nominal threshold voltages of +0.45 V and −0.45 V, respectively. Gaussian Monte Carlo variations are incorporated to represent transistor-parameter uncertainty. The employed model libraries are not calibrated to a specific fabrication process node, such as 180 nm CMOS or 14 nm FinFET. Therefore, the extracted voltage-dependent failure characteristics represent a technology-generic device model rather than a process-specific silicon implementation. These circuits are used to characterize the
impact of supply-voltage scaling and device variations rather than to introduce
new circuit architectures. For each circuit, $V_{\mathrm{DD}}$ was varied over the considered operating
range and Monte Carlo simulations were performed to account for transistor-level
variations. The simulated outputs were post-processed to determine the number of
failed realizations at each supply voltage. The resulting voltage-dependent error
characteristics are subsequently incorporated into the system-level algorithm.

\subsection{SRAM Error Characterization}
\label{subsec:sram_design}

A conventional six-transistor (6T) SRAM bit-cell was implemented in PSpice for memory reliability characterization. The cell consists of two cross-coupled CMOS inverters and two access transistors controlled by the word line (WL) \cite{chao20263}. The internal nodes $Q$ and $\overline{Q}$ store complementary logic values, while BL and BLB represent the corresponding bit lines. The cross-coupled inverters form a bistable storage element that retains one bit of information through regenerative feedback. The SRAM cell was evaluated through Monte Carlo analysis over the considered $V_{\mathrm{DD}}$ range. Device variations affect the relative strengths of the pull-up, pull-down, and access transistors. At higher supply voltages, the cell has sufficient operating margin to tolerate these variations. However, as $V_{\mathrm{DD}}$ approaches the near-threshold region, the transistor overdrive, drive current, and static noise margin decrease. As a result, the stored state becomes more sensitive to device mismatch.

\begin{figure}[t]
    \centering

    \begin{subfigure}[t]{0.90\columnwidth}
        \centering
        \includegraphics[width=0.7\linewidth]{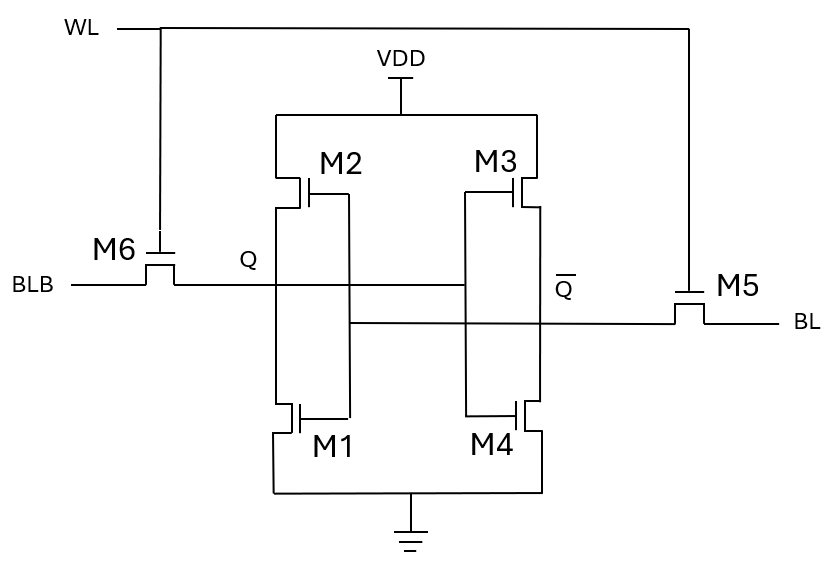}
        \caption{6T SRAM bit-cell.}
        \label{fig:sram_cell}
    \end{subfigure}

    \vspace{0.25cm}

    \begin{subfigure}[t]{0.90\columnwidth}
        \centering
        \includegraphics[width=0.90\linewidth]{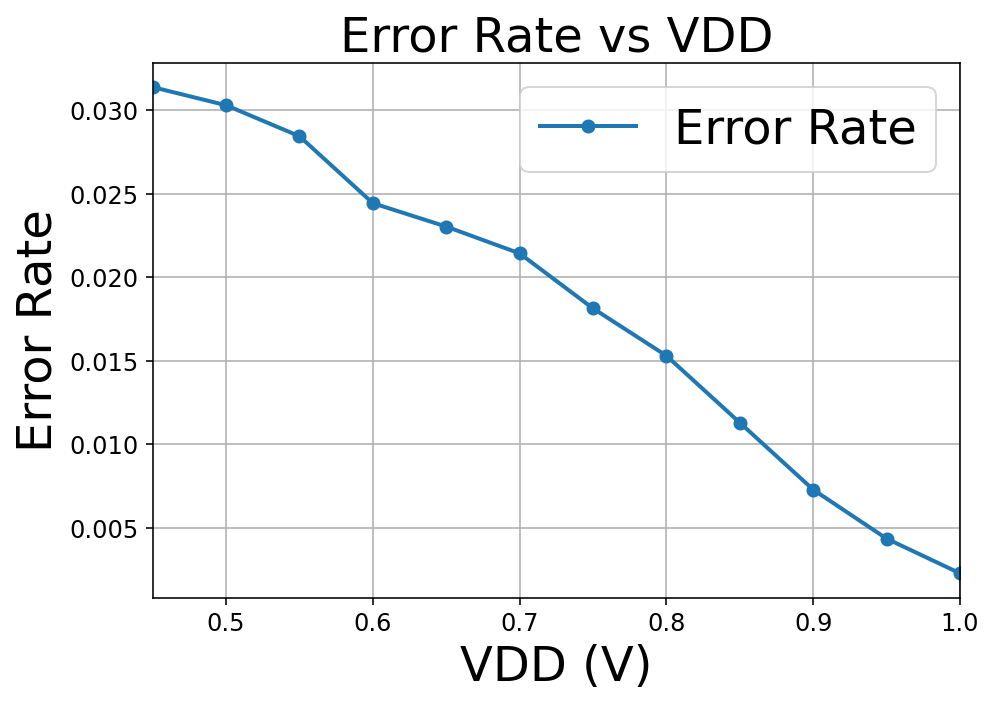}
        \caption{Extracted SRAM error rate as a function of supply voltage.}
        \label{fig:sram_error_rate}
    \end{subfigure}

    \caption{6T SRAM bit-cell and its extracted voltage-dependent error characteristic }
    \label{fig:sram_schematic}
\end{figure}

Variations in transistor parameters can shift the switching points of the cross-coupled inverters and disturb the internal storage nodes. If either $Q$ or $\overline{Q}$ crosses the trip point of the opposite inverter, regenerative feedback can force the cell into an unintended stable state. This condition is classified as an SRAM failure. The PSpice simulation outputs were post-processed using Python. For each $V_{\mathrm{DD}}$, the simulated storage state was compared with the expected state, and realizations that failed to retain the intended logic value were counted as errors. As shown in Fig.~\ref{fig:sram_error_rate}, the SRAM error rate decreases as $V_{\mathrm{DD}}$ increases. The error rate is approximately $3\%$ in the lower-voltage region, decreases to about $1.5\%$ at $V_{\mathrm{DD}}=0.8$~V, falls below $1\%$ near $0.9$~V, and reaches approximately $0.25\%$ at $1.0$~V. This trend demonstrates the increased sensitivity of SRAM operation to device variations at lower supply voltages. The extracted voltage-dependent SRAM error characteristic is subsequently used in the system-level reliability evaluation.

\subsection{Adder Error Characterization}
\label{subsec:adder_design}

An 8-bit ripple-carry adder was implemented in PSpice to characterize
arithmetic reliability under near-threshold-voltage operation. The adder
consists of eight cascaded stages, where the carry output generated by
each stage is supplied to the carry input of the following stage. The
adder produces the sum outputs $S_0$--$S_7$ together with the final carry
output. For reliability characterization, $V_{\mathrm{DD}}$ was swept from
$0.3$~V to $1.0$~V with a step size of $0.01$~V, and 1000 Monte Carlo
runs were performed with device-level and process-level
variations. The sum outputs $S_0$--$S_7$ were evaluated separately for
their corresponding input combinations. The resulting simulation data
were exported from PSpice as CSV files and subsequently processed using
Python.

At each value of $V_{\mathrm{DD}}$, the simulated voltage of the
corresponding sum output was converted into a binary value using
$V_{\mathrm{DD}}/2$ as the logic threshold. The detected output was
then compared with the expected sum bit for the applied input
combination. A Monte Carlo realization producing an incorrect logic
value was classified as a failed run. The failed realizations were
accumulated across the considered input combinations to obtain the
voltage-dependent error characteristic of each sum output. At reduced $V_{\mathrm{DD}}$, the lower transistor overdrive and reduced
logic margin increase the sensitivity of the adder to device variations.
These variations can alter the switching behavior of the internal logic
and produce incorrect sum or carry values. The effect becomes more
significant for higher-order bits because their outputs depend on carry
signals propagated through the preceding stages \cite{safarpour2021algorithm}.

 \begin{figure}[t]
    \centering

    \begin{subfigure}[t]{0.95\columnwidth}
        \centering
        \includegraphics[width=0.95\linewidth]{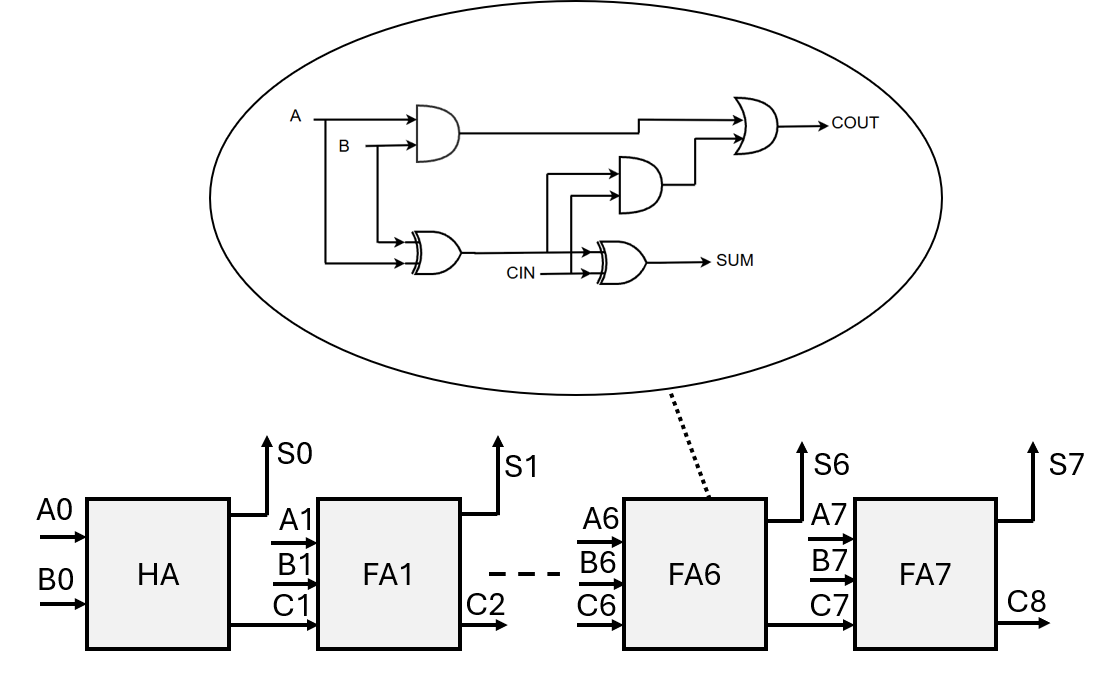}
        \caption{8-bit ripple-carry adder.}
        \label{fig:adder_architecture}
    \end{subfigure}

    \vspace{0.25cm}

    \begin{subfigure}[t]{0.95\columnwidth}
        \centering
        \includegraphics[width=0.90\linewidth]{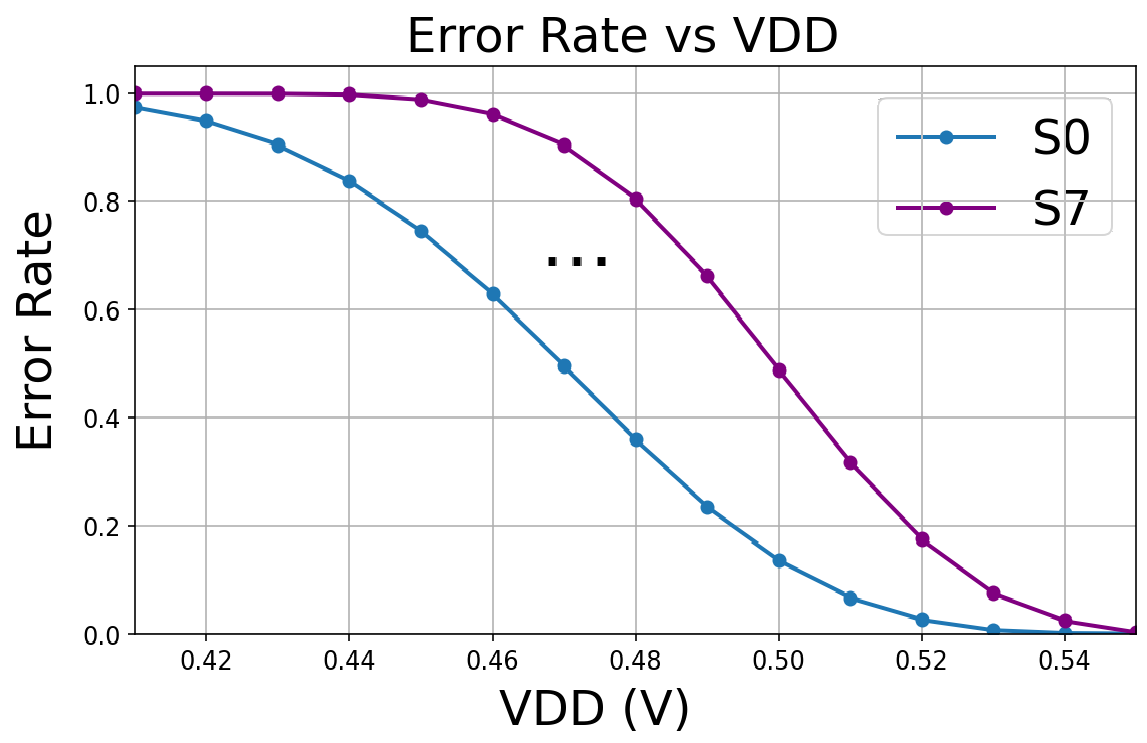}
        \caption{Extracted error rate for $S_0$ and $S_7$.}
        \label{fig:adder_error_rate}
    \end{subfigure}

    \caption{8-bit ripple-carry adder and its voltage-dependent error characteristics.}    \label{fig:adder_characterization}
\end{figure}
 
Fig.~\ref{fig:adder_characterization} presents the implemented 8-bit
ripple-carry adder together with its extracted error characteristics.
As illustrated in the architecture, the least significant stage is
implemented using a half adder (HA), while the remaining stages
$FA_1$--$FA_7$ are implemented using full adders. The carry output of
each stage is propagated to the carry input of the following stage,
resulting in the sequence $C_1$--$C_8$ and the corresponding sum outputs
$S_0$--$S_7$. The inset in Fig.~\ref{fig:adder_architecture} shows the
internal logic of a representative full-adder stage, constructed from
XOR, AND, and OR gates. Each of these logic gates was implemented at the
transistor level in PSpice. So, the Monte Carlo analysis captures
device-level variations within the complete arithmetic structure rather
than relying on ideal logic-gate models. For a compact representation, only $S_0$ and $S_7$ are shown in the
error-rate plot. The intermediate outputs $S_1$--$S_6$ were evaluated
using the same procedure but are omitted for clarity. The $S_0$ output
is generated by the first stage and does not depend on a carry propagated
from a preceding stage, whereas $S_7$ is produced at the final stage and
depends on the carry chain through the preceding full adders. As shown
in Fig.~\ref{fig:adder_error_rate}, the error probability decreases as
$V_{\mathrm{DD}}$ increases. At approximately
$V_{\mathrm{DD}}=0.50$~V, the error rate is about $0.14$ for $S_0$ and
$0.49$ for $S_7$, while both approach zero near
$0.54$--$0.55$~V. The increased sensitivity of $S_7$ can be attributed to the ripple-carry architecture. Errors in intermediate carry signals may propagate through several full-adder stages and influence higher-order sum outputs. At lower $V_{\mathrm{DD}}$, reduced transistor drive strength and smaller logic margins make the gate implementations more sensitive to device variation. As a result, the probability of an incorrect sum or carry increases.
As $V_{\mathrm{DD}}$ increases, the improved operating margins reduce
this sensitivity and the resulting error rate. The extracted
characteristics of $S_0$--$S_7$ are subsequently used to construct the
adder vulnerability profile for the system-level fault model.

 \begin{figure}[t]
    \centering

    \begin{subfigure}[t]{0.95\columnwidth}
        \centering
        \includegraphics[width=0.95\linewidth]{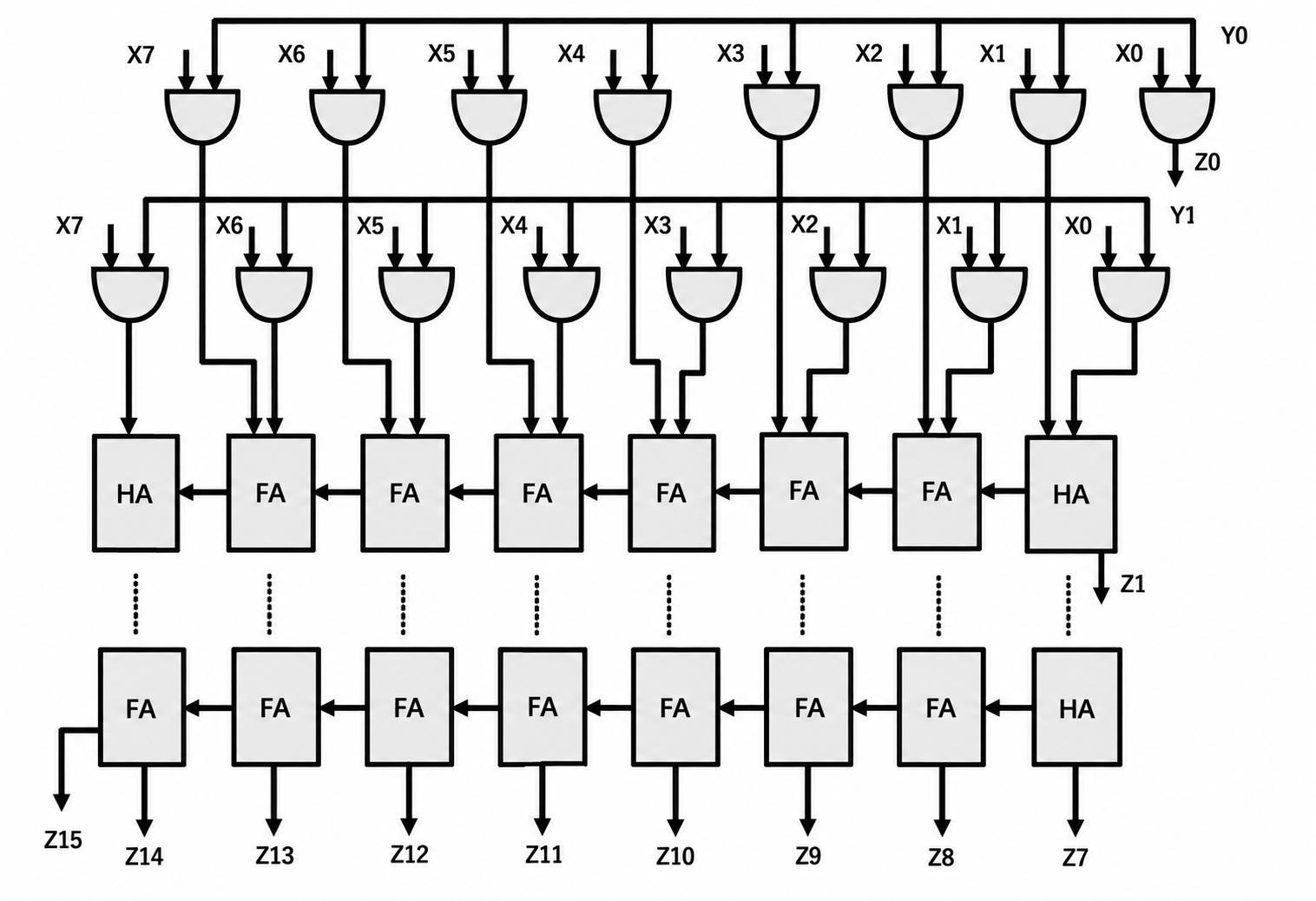}
        \caption{8-bit Wallace-tree multiplier.}
        \label{fig:wallace_architecture}
    \end{subfigure}

    \vspace{0.25cm}

    \begin{subfigure}[t]{0.95\columnwidth}
        \centering
        \includegraphics[width=0.90\linewidth]{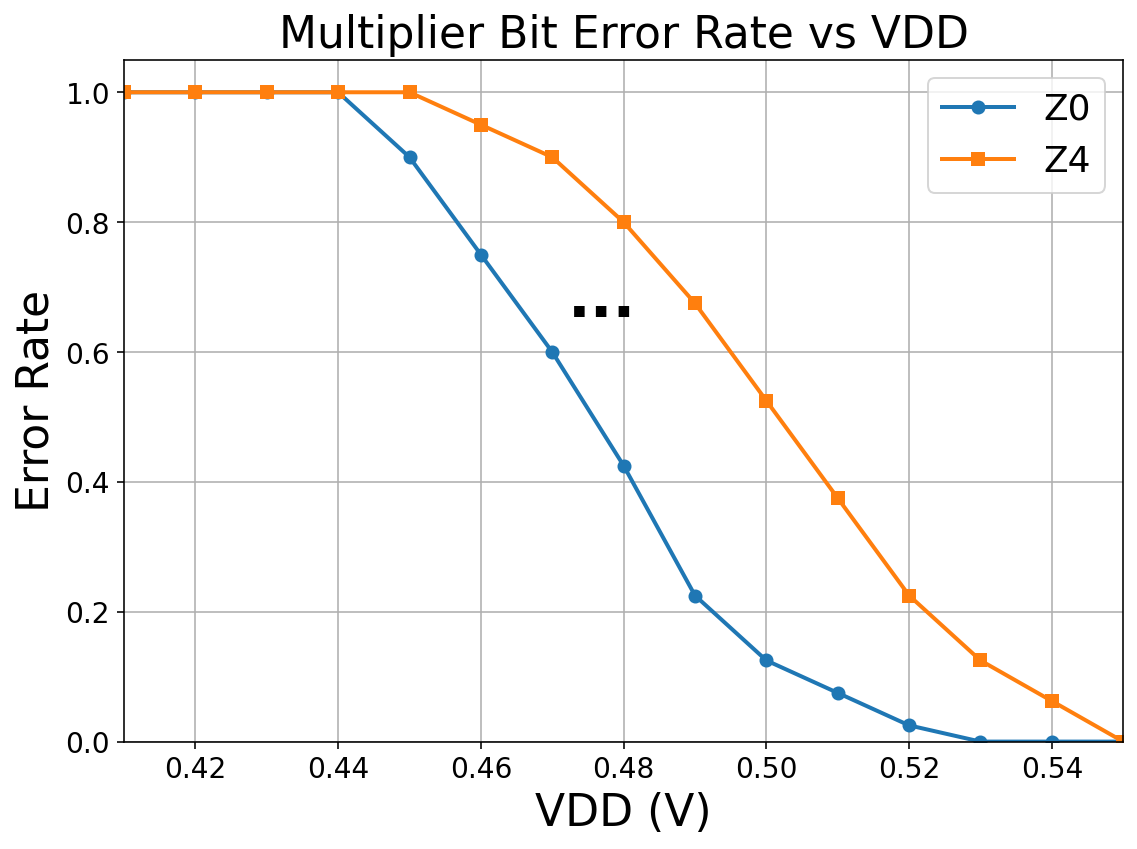}
        \caption{Extracted error rate for $Z_0$ and $Z_4$.}
        \label{fig:multiplier_error_rate}
    \end{subfigure}

    \caption{8-bit Wallace-tree multiplier and its voltage-dependent
    error characteristics.}
    \label{fig:multiplier_characterization}
\end{figure}

\subsection{Multiplier Error Characterization}
\label{subsec:multiplier_design}

An 8-bit Wallace-tree multiplier was implemented in PSpice to characterize
multiplication reliability under near-threshold-voltage operation. The
multiplier generates partial products using AND operations and reduces
them through multiple stages of half adders and full adders to produce
the 16-bit output $Z_0$--$Z_{15}$. The multiplier was evaluated using the same voltage-scaling and Monte
Carlo methodology applied to the adder. For the circuit-level
characterization, the product outputs $Z_0$--$Z_4$ were evaluated
individually for their corresponding input combinations, and the
simulation results were post-processed to obtain the voltage-dependent
error characteristics. At reduced $V_{\mathrm{DD}}$, lower transistor drive strength and reduced
logic margins increase the sensitivity of the partial-product and
reduction stages to device variations. Errors generated at intermediate
nodes can propagate through the Wallace-tree network and affect the final
product bits. Higher-order outputs can therefore exhibit greater
sensitivity because they depend on additional partial products and
reduction stages.

Fig.~\ref{fig:multiplier_characterization} shows the 8-bit Wallace-tree
multiplier together with the extracted error characteristics for $Z_0$
and $Z_4$. The intermediate measured outputs $Z_1$--$Z_3$ were evaluated
using the same procedure but are omitted from the plot for clarity. As shown in Fig.~\ref{fig:multiplier_error_rate}, the error rate decreases
as $V_{\mathrm{DD}}$ increases. At approximately $0.50$~V, the error rate
is about $0.12$ for $Z_0$ and $0.70$ for $Z_4$. The $Z_0$ output
approaches zero near $0.54$~V, whereas $Z_4$ remains more sensitive over
the same voltage range. This difference reflects the greater number of
intermediate reduction operations contributing to $Z_4$. The extracted characteristics of $Z_0$--$Z_4$ are subsequently used to
construct the multiplier error model for the system-level evaluation.

\subsection{Circuit-Derived Error Model Construction}
\label{subsec:error_model_construction}

The circuit-level characterization presented in the preceding subsections
provides measured error information for the SRAM, adder, and multiplier.
However, the system-level massive MIMO implementation uses a 16-bit
datapath. Therefore, the measured circuit-level results were converted
into 16-bit vulnerability profiles for use during fault injection. The
measured values were retained directly where available, while linear
regression was used to estimate the unmeasured bit positions of the
adder and multiplier.

\vspace{4pt}

\subsubsection{SRAM Error Model}

For the SRAM model, the measured memory error characteristic was applied
uniformly across the 16-bit word. Each memory-bit position
$M_0$--$M_{15}$ was therefore assigned a vulnerability weight of $0.03$.
No regression or bit-position extrapolation was applied to the SRAM
model.

\subsubsection{Adder Error Model}

Circuit-level characterization provided measured vulnerability values
for the eight adder outputs $S_0$--$S_7$. Since the system-level
datapath contains 16 bits, the remaining positions $S_8$--$S_{15}$ were
estimated from the measured data using first-order linear regression. The measured values were fitted as a function of bit position $n$ using

\begin{equation}
P_S(n) = a_S n + b_S ,
\end{equation}
where the fitted slope and intercept were obtained as
$a_S=0.03785714$ and $b_S=0.2875$. Thus, the regression model is

\begin{equation}
P_S(n) = 0.03785714n + 0.2875.
\label{eq:adder_regression}
\end{equation}
The measured values for $S_0$--$S_7$ were retained directly, while
Eq.~\ref{eq:adder_regression} was used only to estimate
$S_8$--$S_{15}$. The resulting 16-bit adder vulnerability profile is
listed in Table~\ref{tab:adder_vulnerability}. The increasing trend reflects the greater vulnerability assigned to
higher-order adder bits. The complete set of measured and extrapolated
values is subsequently used as the adder bit-vulnerability profile in
the system-level fault model.

\begin{table}[t]
\centering
\caption{Measured and predicted 16-bit adder vulnerability weights.}
\label{tab:adder_vulnerability}
\footnotesize
\begin{tabular}{c c c | c c c}
\hline
Bit & Error Susceptibility & Source & Bit & Weight & Source \\
\hline
$S_0$  & 0.150000 & Measured  & $S_8$  & 0.590357 & Predicted \\
$S_1$  & 0.400000 & Measured  & $S_9$  & 0.628214 & Predicted \\
$S_2$  & 0.420000 & Measured  & $S_{10}$ & 0.666071 & Predicted \\
$S_3$  & 0.450000 & Measured  & $S_{11}$ & 0.703929 & Predicted \\
$S_4$  & 0.460000 & Measured  & $S_{12}$ & 0.741786 & Predicted \\
$S_5$  & 0.470000 & Measured  & $S_{13}$ & 0.779643 & Predicted \\
$S_6$  & 0.500000 & Measured  & $S_{14}$ & 0.817500 & Predicted \\
$S_7$  & 0.510000 & Measured  & $S_{15}$ & 0.855357 & Predicted \\
\hline
\end{tabular}
\end{table}

\subsubsection{Multiplier Error Model}

For the multiplier, circuit-level measurements were available for the
product outputs $Z_0$--$Z_4$. A linear regression model was therefore
constructed from these measured values to estimate the remaining
positions $Z_5$--$Z_{15}$. The fitted relationship is expressed as

\begin{equation}
P_Z(n) = a_Z n + b_Z ,
\end{equation}
where $a_Z=0.134$ and $b_Z=0.232$. The resulting regression model is

\begin{equation}
P_Z(n) = 0.134n + 0.232.
\label{eq:multiplier_regression}
\end{equation}
The measured values for $Z_0$--$Z_4$ were preserved, while
Eq.~\ref{eq:multiplier_regression} was used to predict
$Z_5$--$Z_{15}$. Since the vulnerability weights are constrained to the
range $[0,1]$, predicted values exceeding unity were clipped to $1.0$.
The resulting 16-bit multiplier vulnerability profile is shown in
Table~\ref{tab:multiplier_vulnerability}. The resulting SRAM, adder, and multiplier profiles are used as relative
bit-vulnerability weights during fault injection in the subsequent
system-level simulations. The measured values therefore preserve the
circuit-level reliability information, while the regression-based
estimates extend the available characterization to the complete 16-bit
datapath.

\begin{table}[t]
\centering
\caption{Measured and predicted 16-bit multiplier vulnerability weights.}
\label{tab:multiplier_vulnerability}
\footnotesize
\begin{tabular}{c c c | c c c}
\hline
Bit & Error Susceptibility & Source & Bit & Weight & Source \\
\hline
$Z_0$  & 0.160000 & Measured  & $Z_8$  & 1.000000 & Predicted \\
$Z_1$  & 0.380000 & Measured  & $Z_9$  & 1.000000 & Predicted \\
$Z_2$  & 0.620000 & Measured  & $Z_{10}$ & 1.000000 & Predicted \\
$Z_3$  & 0.640000 & Measured  & $Z_{11}$ & 1.000000 & Predicted \\
$Z_4$  & 0.700000 & Measured  & $Z_{12}$ & 1.000000 & Predicted \\
$Z_5$  & 0.902000 & Predicted & $Z_{13}$ & 1.000000 & Predicted \\
$Z_6$  & 1.000000 & Predicted & $Z_{14}$ & 1.000000 & Predicted \\
$Z_7$  & 1.000000 & Predicted & $Z_{15}$ & 1.000000 & Predicted \\
\hline
\end{tabular}
\end{table}

\begin{figure}[h]
\centering
\includegraphics[width=0.92\columnwidth]{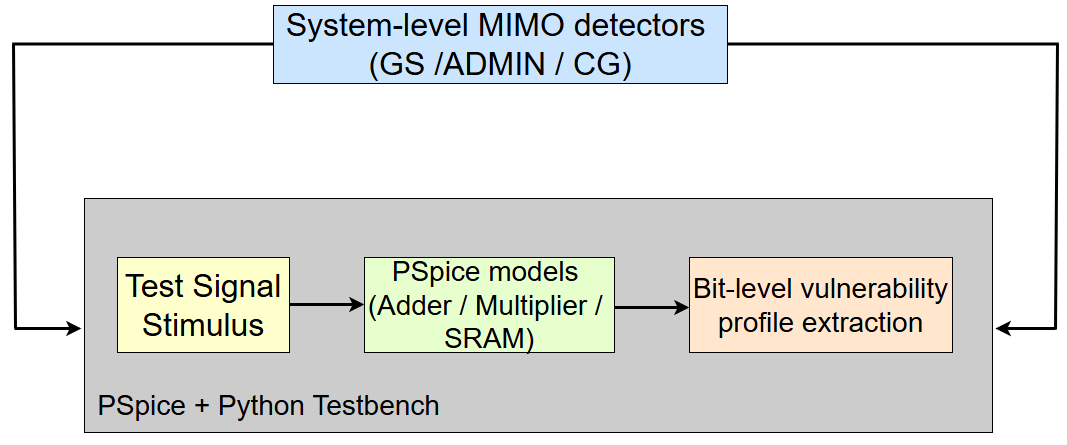}
\caption{PSPICE Python testbench for voltage-dependent fault injection in massive MIMO detectors.}
\label{fig}
\end{figure}

The circuit-level characterization is integrated with the system-level MIMO simulation through the cross-layer testbench shown in Fig.~\ref{fig}. Test stimuli are first applied to the PSpice models of the adder, multiplier, and SRAM over the considered supply-voltage range. The resulting circuit responses are processed to extract voltage-dependent bit-level vulnerability profiles. These profiles are then incorporated into the Python-based fault-injection framework and mapped to the corresponding arithmetic and memory operations of the GS, ADMIN, and CG detectors. In this way, the hardware faults introduced during the MIMO simulations are derived from the circuit-level characterization rather than from a uniform random fault model. The resulting framework provides the connection between transistor-level reliability behavior and the system-level impact of computational errors on massive MIMO detection.

\section{Algorithm Based Fault Tolerance Algorithm}
\label{sec:abft_tmr}

The complexity analysis identifies multiplication and addition as the dominant computational operations of the considered detectors. Accordingly, the circuit derived fault characteristics are mapped to their corresponding arithmetic blocks. SRAM reliability is modeled separately using its measured voltage dependent failure profile. The resulting PSpice derived error probabilities are incorporated into a Python based fault injection framework for GS, ADMIN, and CG. During each arithmetic fault event, a result word and bit position are selected based on the corresponding circuit derived vulnerability profile. The selected bit is then flipped. Because fault events can occur repeatedly across the many arithmetic operations of a detector execution, multiple bit corruptions may accumulate during a MIMO trial. Detector specific operation-count scaling is also included to reflect the different arithmetic workloads of GS, ADMIN, and CG. ABFT detects computational errors by propagating checksum information through the detector. At the final output, the propagated checksum is compared with the checksum recomputed from the recovered data to identify potential errors.

\subsection{Checksum-Based ABFT Error Detection}
\label{subsec:abft_detection}

ABFT detects computational errors by
exploiting checksum properties preserved by linear-algebra operations. Following the conventional formulation in
\cite{rinkinen2025near,aliagha2025scissors}, for an
$N\times N$ matrix $\mathbf{A}$ and an all-one vector
$\mathbf{e}=[1,1,\ldots,1]$, the row-checksum matrix is

\begin{equation}
\mathbf{A}^{r}
=
\left[
\begin{array}{cc}
\mathbf{A} & \mathbf{A}\mathbf{e}^{T}
\end{array}
\right],
\label{eq:abft_row_checksum}
\end{equation}
where the appended column contains the row sums of $\mathbf{A}$. The
corresponding column-checksum matrix is

\begin{equation}
\mathbf{A}^{c}
=
\left[
\begin{array}{c}
\mathbf{A}\\
\mathbf{e}\mathbf{A}
\end{array}
\right].
\label{eq:abft_column_checksum}
\end{equation}
For matrix multiplication $\mathbf{C}=\mathbf{A}\mathbf{B}$, the
checksum property is preserved as

\begin{equation}
\mathbf{C}^{f}
=
\mathbf{A}^{c}\mathbf{B}^{r}
=
\left[
\begin{array}{cc}
\mathbf{A}\mathbf{B} &
\mathbf{A}\mathbf{B}\mathbf{e}^{T}\\
\mathbf{e}\mathbf{A}\mathbf{B} &
\mathbf{e}\mathbf{A}\mathbf{B}\mathbf{e}^{T}
\end{array}
\right].
\label{eq:abft_matrix_mult}
\end{equation}
A computational fault can therefore be detected when the checksum
associated with the computed data is inconsistent with the expected
checksum relation. These equations provide the conventional
checksum-preserving principle on which the ABFT mechanism used in this
work is based. In the present implementation, checksum information is incorporated directly into the detector state. For a detector state vector \(\mathbf{x}=[x_1,x_2,\ldots,x_N]\), the checksum is defined as \(x_{\mathrm{chk}}=\sum_{i=1}^{N}x_i\), and the checksum-augmented state is represented as \(\mathbf{x}_{\mathrm{ABFT}}=[x_1,x_2,\ldots,x_N,x_{\mathrm{chk}}]\). To describe checksum propagation, we define the checksum discrepancy as \(\delta_{\mathbf{x}}=x_{\mathrm{chk}}-\mathbf{e}^{T}\mathbf{x}\). For a general transformation \(\mathbf{y}=f(\mathbf{x})\), the carried checksum can be expressed as \(y_{\mathrm{chk}}=\mathbf{e}^{T}\mathbf{y}+T_f(\delta_{\mathbf{x}})\), where \(T_f(\cdot)\) specifies how the incoming discrepancy is propagated. Therefore, the checksum mechanism does not require the generally nonlinear mapping \(f(\cdot)\) itself to preserve the vector sum. For the linear arithmetic used in the detectors, checksum discrepancies propagate naturally through addition and scalar multiplication, such that \(\delta_{\mathbf{x}+\mathbf{y}}=\delta_{\mathbf{x}}+\delta_{\mathbf{y}}\) and \(\delta_{\alpha\mathbf{x}}=\alpha\delta_{\mathbf{x}}\). These relations are used to propagate checksum information through the linear operations appearing in all three detectors.
\begin{figure}[t]
    \centering
    \includegraphics[width=0.95\columnwidth]{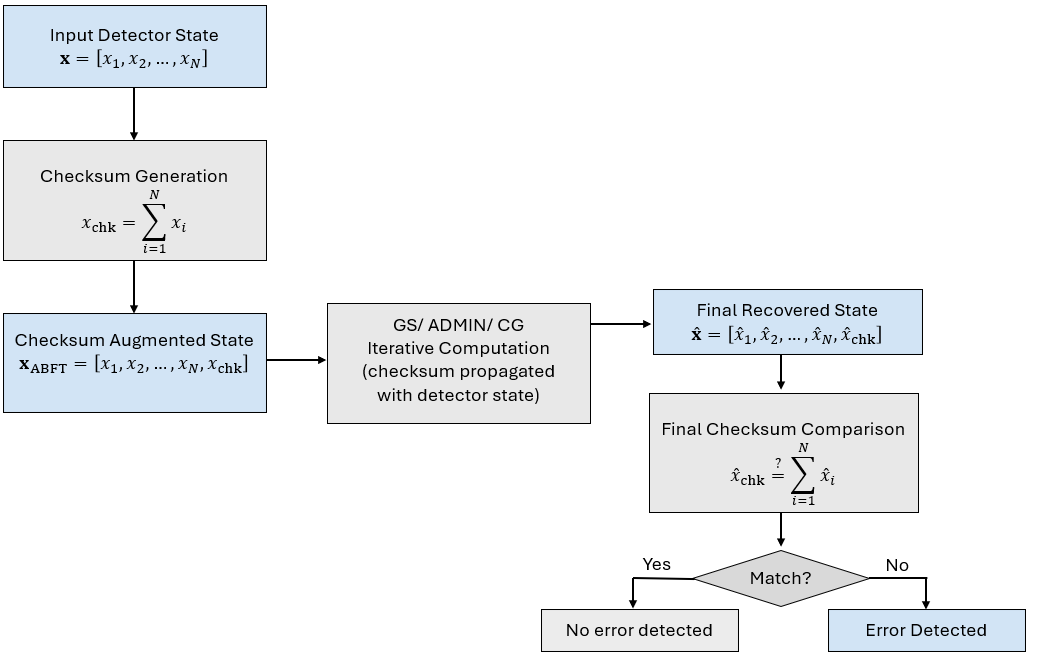}
    \caption{Checksum-based ABFT error-detection flow. The arithmetic
    result generates the actual checksum, while the operation inputs
    independently generate the expected checksum. A mismatch activates
    the recovery mechanism.}
    \label{fig:abft_checksum_flow}
\end{figure}
For GS, the update \(\mathbf{x}^{(k+1)}=(\mathbf{D}+\mathbf{L})^{-1}(\mathbf{b}-\mathbf{U}\mathbf{x}^{(k)})\) is implemented using matrix--vector multiplication, vector addition, and a precomputed linear solve operator. The checksum discrepancy is propagated through each of these linear stages according to the relations above. Consequently, if the input state and matched-filter vector are checksum-consistent, the resulting GS state also remains checksum-consistent during fault-free operation. For ADMIN, the same discrepancy-propagation principle is applied to its linear solves and subsequent primal and dual-variable updates. For the nonlinear projection operation, the projected vector is formed first and its checksum \(\mathbf{e}^{T}\mathbf{u}\) is recomputed directly, so the projection itself is not required to preserve the incoming vector sum. For CG, the checksum discrepancy is propagated through the recursive vector updates using the same linear relations. For example, the solution update propagates as \(\delta_{\mathbf{x}^{(k+1)}}=\delta_{\mathbf{x}^{(k)}}+\alpha_k\delta_{\mathbf{p}^{(k)}}\), while the search-direction update propagates as \(\delta_{\mathbf{p}^{(k+1)}}=\delta_{\mathbf{r}^{(k+1)}}+\beta_k\delta_{\mathbf{p}^{(k)}}\). The CG divisions used to obtain \(\alpha_k\) and \(\beta_k\) produce scalar coefficients rather than checksum-bearing vectors. In the present fault model, the divider is treated as reliable, and the resulting scalar coefficients are applied consistently to both the protected vector data and the carried discrepancy. Therefore, the checksum protects the subsequent vector arithmetic but does not independently validate the scalar coefficient computation. After the final detector iteration, only one ABFT decision is performed. For the final recovered augmented vector \(\widetilde{\mathbf{x}}=[\widehat{\mathbf{x}},\widehat{x}_{\mathrm{chk}}]\), an error is detected when \(\widehat{x}_{\mathrm{chk}}\neq\mathbf{e}^{T}\widehat{\mathbf{x}}\). Thus, detector-specific intermediate transformations maintain checksum consistency without requiring ABFT decisions at intermediate stages. The final comparison then serves as the sole error-detection decision in this work. The comparison is performed
using the same signed 16-bit fixed-point representation as the detector
datapath. Algorithm~\ref{alg:final_abft} summarizes the final-output checksum-based ABFT procedure used in this work.

\begin{algorithm}[t]
\caption{Final Checksum-Based ABFT Detection}
\label{alg:final_abft}
\begin{algorithmic}[1]

\REQUIRE Detector inputs and initial state vector
$\mathbf{x}=[x_1,x_2,\ldots,x_N]$

\ENSURE Final recovered vector $\hat{\mathbf{x}}$
and ABFT detection flag

\STATE $x_{\mathrm{chk}} \leftarrow
\sum_{i=1}^{N}x_i$

\STATE $\mathbf{x}_{\mathrm{ABFT}} \leftarrow
[x_1,x_2,\ldots,x_N,x_{\mathrm{chk}}]$

\FOR{each detector iteration}

    \STATE Update $\mathbf{x}_{\mathrm{ABFT}}$
    according to GS, ADMIN, or CG

    \STATE Propagate the checksum information
    with the detector state

    \STATE Apply the voltage-dependent fault model
    to the corresponding operation

\ENDFOR

\STATE Obtain final augmented output
$\hat{\mathbf{x}}_{\mathrm{ABFT}}$

\STATE $\hat{\mathbf{x}} \leftarrow
\hat{\mathbf{x}}_{\mathrm{ABFT}}(1:N)$

\STATE $\hat{x}_{\mathrm{chk}} \leftarrow
\hat{\mathbf{x}}_{\mathrm{ABFT}}(N+1)$

\STATE $\hat{x}_{\mathrm{sum}} \leftarrow
\sum_{i=1}^{N}\hat{x}_i$

\IF{$\hat{x}_{\mathrm{chk}} \neq
\hat{x}_{\mathrm{sum}}$}

    \STATE $\mathrm{ABFT\_flag} \leftarrow 1$
    \COMMENT{Error detected}

\ELSE

    \STATE $\mathrm{ABFT\_flag} \leftarrow 0$
    \COMMENT{No error detected}

\ENDIF

\STATE \textbf{return}
$\hat{\mathbf{x}},\mathrm{ABFT\_flag}$

\end{algorithmic}
\end{algorithm}

\subsection{Final-Output Error Classification}
\label{subsec:final_error_classification}

The effectiveness of the checksum-based ABFT mechanism is evaluated at
the final recovered detector output. An actual computational error is
recorded when the final faulty recovered vector differs from the
corresponding fault-free fixed-point output. If an actual error also
produces a checksum mismatch, it is classified as an ABFT-detected
error. If the recovered vector is corrupted while the final checksum
relation remains valid, the error is classified as a silent or
undetected error. The final-output error counts therefore satisfy

\begin{equation}
N_{\mathrm{actual}}
=
N_{\mathrm{detected}}
+
N_{\mathrm{silent}}.
\label{eq:error_classification}
\end{equation}
Here, $N_{\mathrm{actual}}$ denotes the total number of trials in which the final faulty recovered vector differs from the corresponding fault-free fixed-point output, $N_{\mathrm{detected}}$ denotes the number of those errors identified by a final checksum mismatch, and $N_{\mathrm{silent}}$ denotes the number of erroneous final outputs that pass the checksum verification. These metrics are used to evaluate the detection behavior of GS, ADMIN,
and CG across the considered supply-voltage range.

\section{Results and Discussion}
\label{sec:results}

This section evaluates the impact of voltage-dependent hardware faults on the GS, ADMIN, and CG massive MIMO detectors. It also examines the effectiveness of the proposed checksum-based ABFT mechanism. The analysis focuses on the onset of computational errors near the point of first failure (PoFF) and the resulting BER degradation. It also examines the detection behavior of ABFT. Furthermore, the analysis investigates how detector workloads and error-propagation characteristics influence its effectiveness.

\subsection{Simulation Setup}
\label{subsec:simulation_setup}

Unless otherwise stated, the system-level simulations consider a
$64\times16$ massive MIMO configuration with 64-QAM modulation and a
communication Signal-to-Noise Ratio (SNR) of $15$~dB. The GS, ADMIN, and CG detectors are evaluated using three iterative updates. All three detectors employ a signed 16-bit fixed-point representation with nine fractional bits. For each supply-voltage point, 1000 independent
MIMO trials are performed. The circuit-derived voltage-dependent fault profiles obtained from the PSpice
characterization are applied to the corresponding arithmetic operations during
detector execution. The effective fault exposure is scaled according to the
operation counts of each detector. The main PoFF analysis considers arithmetic
faults in the adder and multiplier, while SRAM faults are evaluated separately
using the measured SRAM failure profile. To obtain smoother BER estimates in the region surrounding the observed PoFF,
the near-PoFF simulations are repeated over five independent runs and the
reported BER values are averaged across the runs. The same trials are also
used to determine the rate of actual computational errors at the final recovered
detector output.

\subsection{Final-Output Error Detection and PoFF Behavior}
\label{subsec:poff_detection}

\begin{figure*}[t]
    \centering

    \begin{subfigure}[t]{0.32\textwidth}
        \centering
        \includegraphics[width=\linewidth]{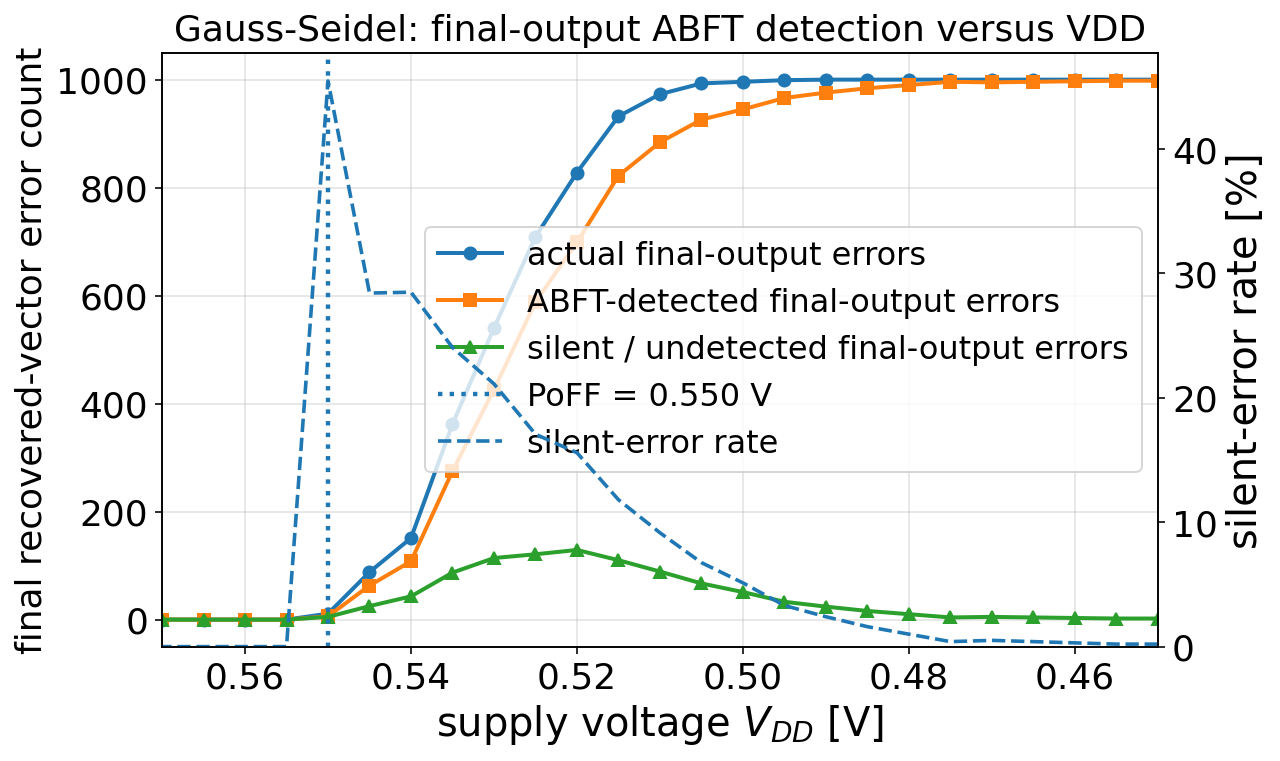}
        \caption{Gauss--Seidel}
        \label{fig:gs_poff_detection}
    \end{subfigure}
    \hfill
    \begin{subfigure}[t]{0.32\textwidth}
        \centering
        \includegraphics[width=\linewidth]{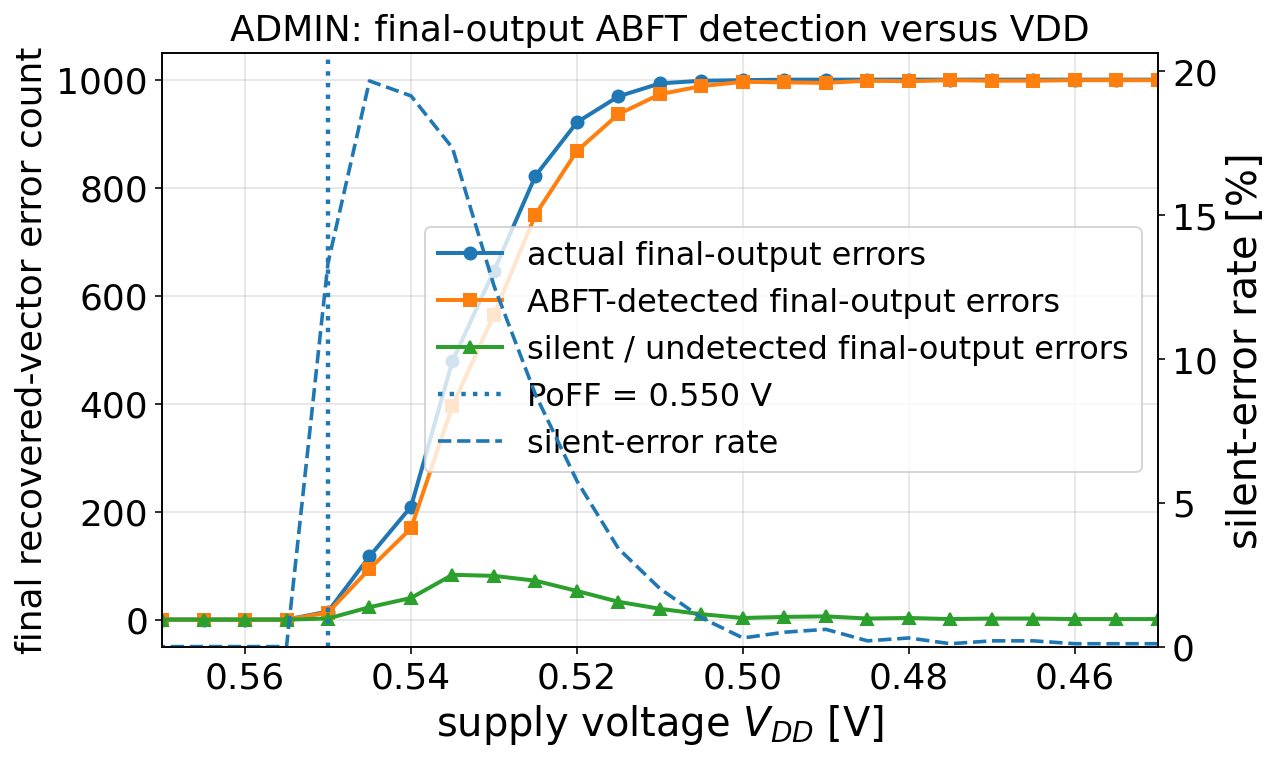}
        \caption{ADMIN}
        \label{fig:admin_poff_detection}
    \end{subfigure}
    \hfill
    \begin{subfigure}[t]{0.32\textwidth}
        \centering
        \includegraphics[width=\linewidth]{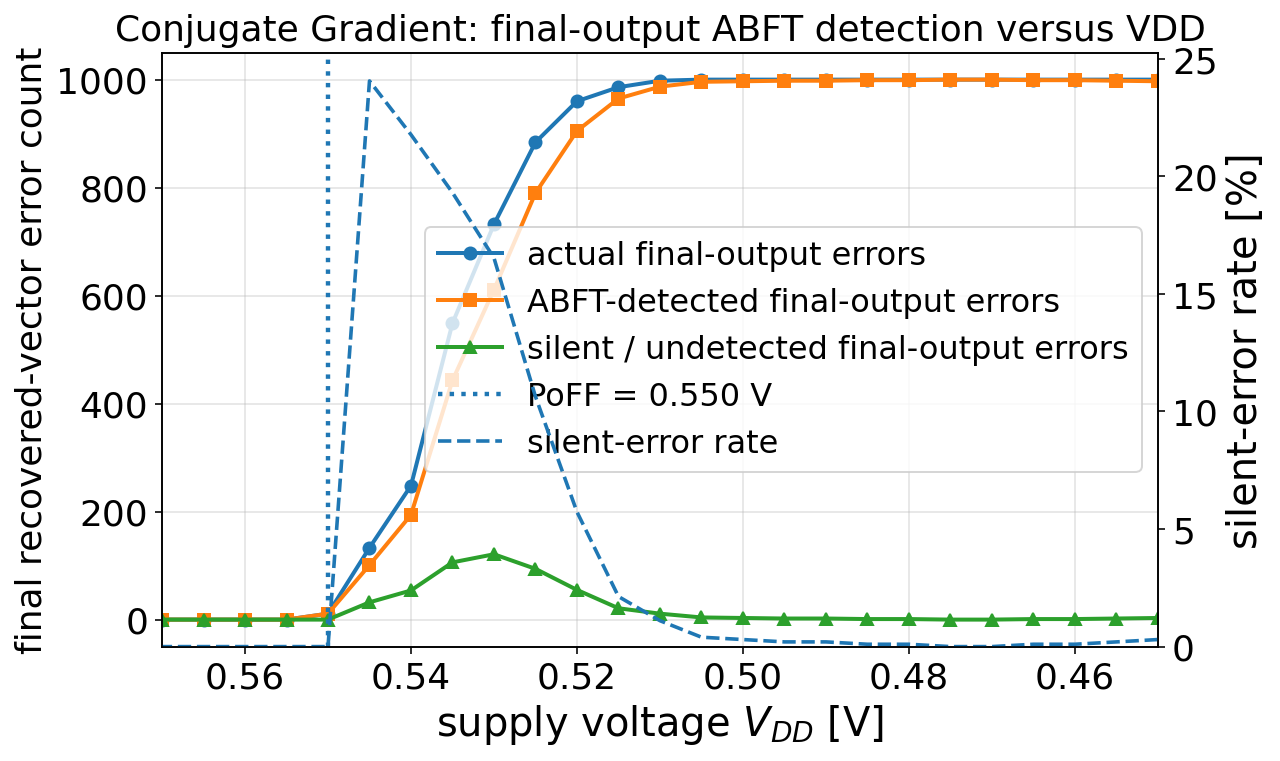}
        \caption{Conjugate Gradient}
        \label{fig:cg_poff_detection}
    \end{subfigure}

    \caption{Final-output computational errors and checksum-based ABFT
detection versus supply voltage for (a) GS, (b) ADMIN, and (c) CG.
The vertical dotted line indicates the observed arithmetic PoFF at
$V_{DD}=0.550$~V.}

    \label{fig:poff_detection}
\end{figure*}

Fig.~\ref{fig:poff_detection} shows the final-output computational-error
behavior of GS, ADMIN, and CG as the supply voltage is reduced. For each detector, the actual-error count represents the number of trials with an incorrect final recovered vector. These errors are identified by comparing the faulty output with the corresponding fault-free fixed-point result. The detected-error count represents the errors identified by the final ABFT checksum comparison. Errors that remain undetected are classified as silent errors. For all three detectors, essentially no final-output computational errors
are observed at the higher supply voltages. The first observed errors occur at approximately $V_{DD}=0.550$~V. This voltage marks the PoFF in the arithmetic fault study. As the supply voltage decreases below the PoFF, the number of actual final-output errors increases rapidly. Eventually, these errors occur in nearly all simulated trials. The ABFT-detected error count closely follows the actual-error count as
the voltage is reduced. Near the onset of failure, a small fraction of corrupted outputs may still satisfy the final checksum relation. As a result, these errors remain undetected. This produces the observed silent-error component.
The silent-error count initially increases as computational faults become
more frequent, but subsequently decreases as the operating voltage is
reduced further. Consequently, the gap between the actual and
ABFT-detected error counts becomes small in the deeper undervolting
region. The three detectors exhibit different detailed error trajectories even
though they are evaluated using the same circuit-derived voltage-dependent
fault characteristics. This difference results from their distinct
arithmetic workloads and iterative error-propagation behavior. The
results therefore show that the impact of a given circuit-level fault
environment depends not only on the underlying hardware fault
probability, but also on how faults propagate through the detector
algorithm.

\subsection{BER Robustness Near PoFF}
\label{subsec:ber_poff}

\begin{figure*}[t]
    \centering

    \begin{subfigure}[t]{0.32\textwidth}
        \centering
        \includegraphics[width=\linewidth]{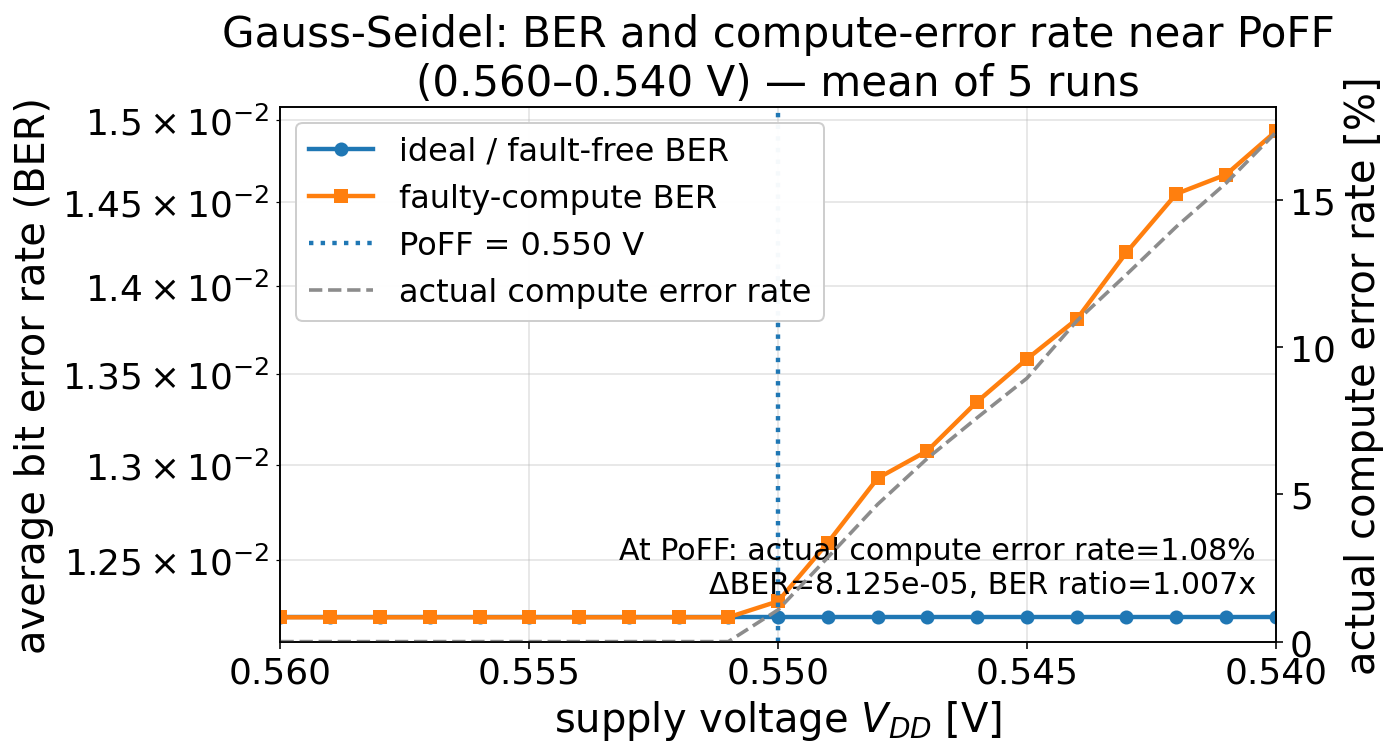}
        \caption{Gauss--Seidel}
        \label{fig:gs_ber_poff}
    \end{subfigure}
    \hfill
    \begin{subfigure}[t]{0.32\textwidth}
        \centering
        \includegraphics[width=\linewidth]{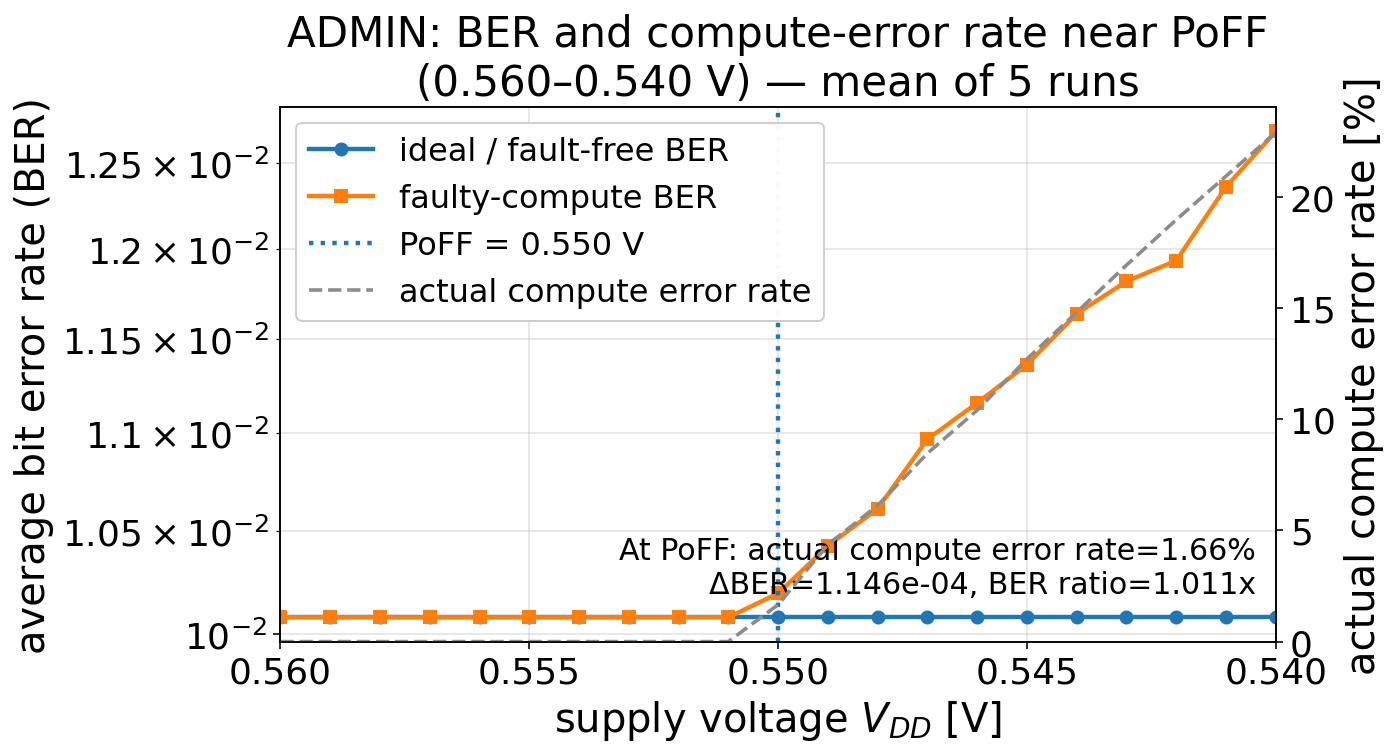}
        \caption{ADMIN}
        \label{fig:admin_ber_poff}
    \end{subfigure}
    \hfill
    \begin{subfigure}[t]{0.32\textwidth}
        \centering
        \includegraphics[width=\linewidth]{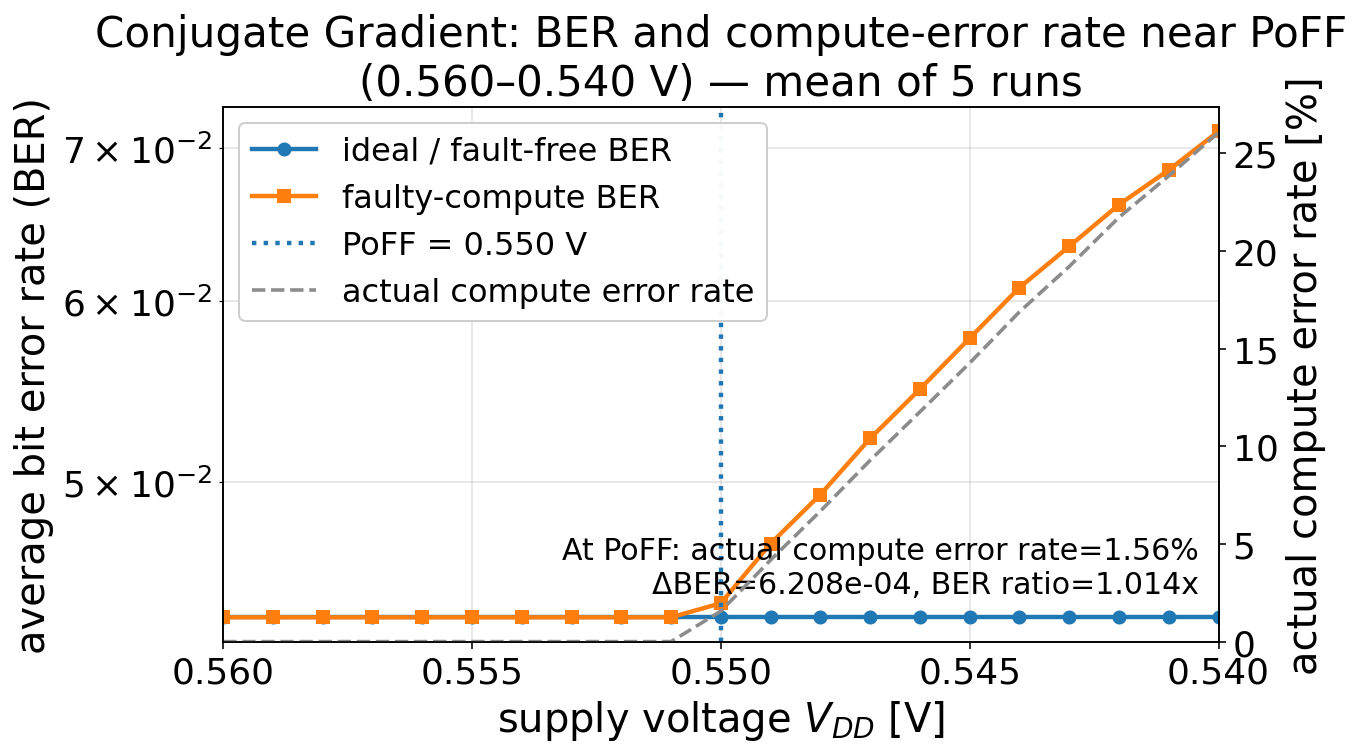}
        \caption{Conjugate Gradient}
        \label{fig:cg_ber_poff}
    \end{subfigure}

    \caption{Average BER and actual compute-error rate near the observed
    PoFF for (a) GS, (b) ADMIN, and (c) CG. BER values are averaged over
    five independent simulation runs. The dashed curve on the secondary
    axis represents the actual compute-error rate.}

    \label{fig:ber_near_poff}
\end{figure*}

The behavior near PoFF is examined in greater detail in
Fig.~\ref{fig:ber_near_poff}. To reduce Monte Carlo variation, the BER
results are averaged over five independent simulation runs. The actual compute-error rate is shown on the secondary axis. It represents the fraction of MIMO trials in which the final faulty recovered vector differs from the corresponding fault-free output. As shown in Fig.~\ref{fig:ber_near_poff}, the actual compute-error rate begins to increase at the observed PoFF. However, the faulty-compute BER initially remains close to the fault-free BER. As $V_{DD}$ is reduced further below PoFF, the compute-error rate
increases more rapidly. Consequently, the difference between the faulty-compute and fault-free BER curves becomes more pronounced. At $V_{DD}=0.550$~V, the observed actual compute-error rates are
approximately 1.08\% for GS, 1.66\% for ADMIN, and 1.56\% for CG. At the
same operating point, the BER increase relative to the fault-free baseline
remains small. The BER ratios are approximately $1.007\times$ for GS,
$1.011\times$ for ADMIN, and $1.014\times$ for CG. Among the three detectors, GS exhibits the highest tolerance to circuit-derived arithmetic faults near PoFF, followed by ADMIN. In contrast, CG experiences the largest BER degradation. This ordering cannot be explained by arithmetic workload alone, since GS has the largest arithmetic workload among the three detectors. Instead, the algorithmic error-propagation structure plays an important role. The sequential Gauss--Seidel update repeatedly refines the solution using the most recently available estimates. This process can reduce the impact of small numerical perturbations on the final hard decision. ADMIN propagates numerical perturbations through its coupled optimization-variable updates. As a result, it exhibits an intermediate level of sensitivity to arithmetic faults. In contrast, CG recursively updates its residual and search direction. This allows arithmetic perturbations to propagate through subsequent search directions and solution updates across iterations. This deeper computational dependency helps explain the larger BER degradation observed for CG. Because the BER remains nearly unchanged when only a small fraction of computational errors occurs. Thus, the decision to retain the detected result or repeat MIMO detection is determined by the desired level of reliability. Recomputation incurs the cost of one additional detector execution. In either case, the clock setting is adjusted first. The supply voltage is then increased by one voltage step to move the system toward a more reliable operating point. 

\subsection{SRAM Reliability and ABFT Detection}
\label{subsec:sram_results}

\begin{figure*}[t]
    \centering

    \begin{subfigure}[t]{0.32\textwidth}
        \centering
        \includegraphics[width=\linewidth]{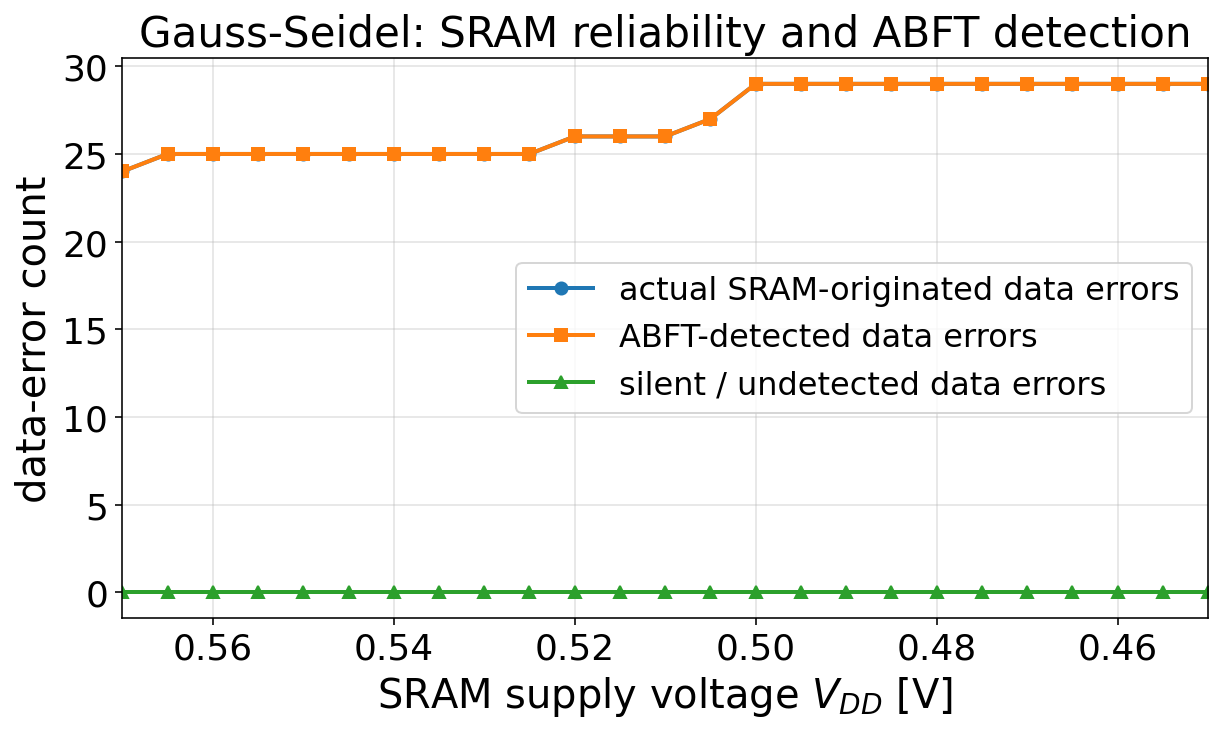}
        \caption{Gauss--Seidel}
        \label{fig:gs_sram_detection}
    \end{subfigure}
    \hfill
    \begin{subfigure}[t]{0.32\textwidth}
        \centering
        \includegraphics[width=\linewidth]{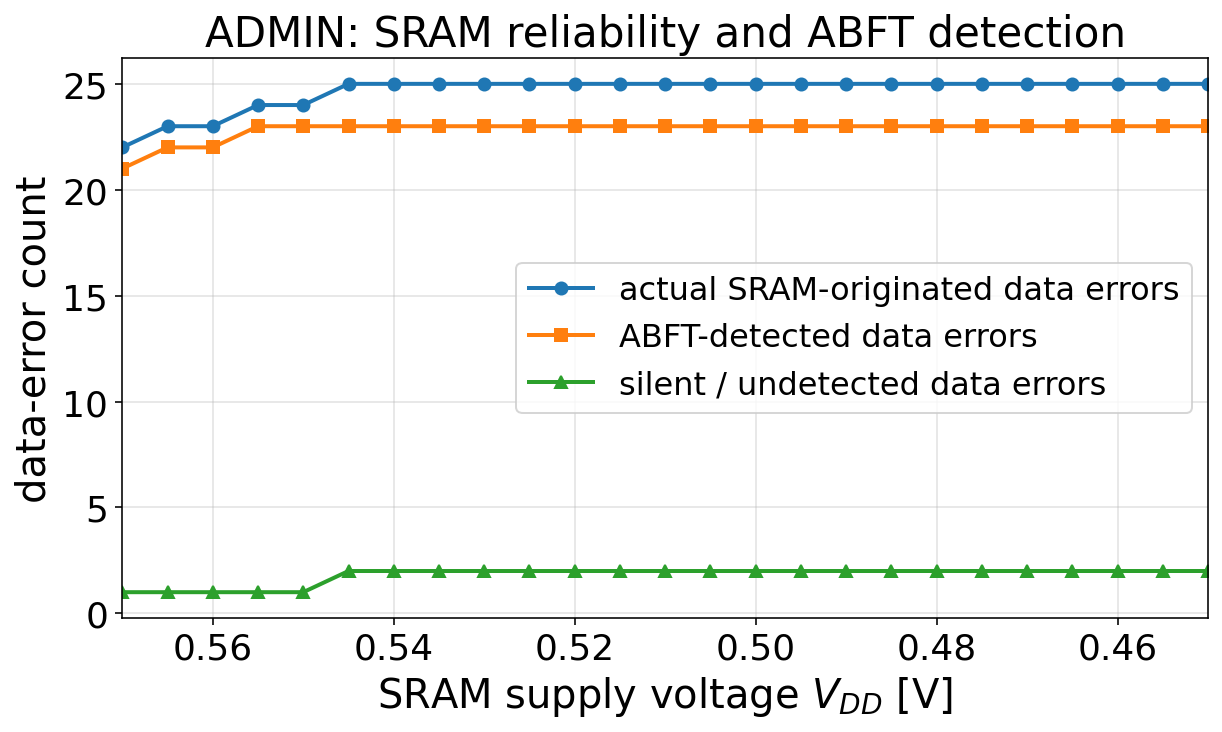}
        \caption{ADMIN}
        \label{fig:admin_sram_detection}
    \end{subfigure}
    \hfill
    \begin{subfigure}[t]{0.32\textwidth}
        \centering
        \includegraphics[width=\linewidth]{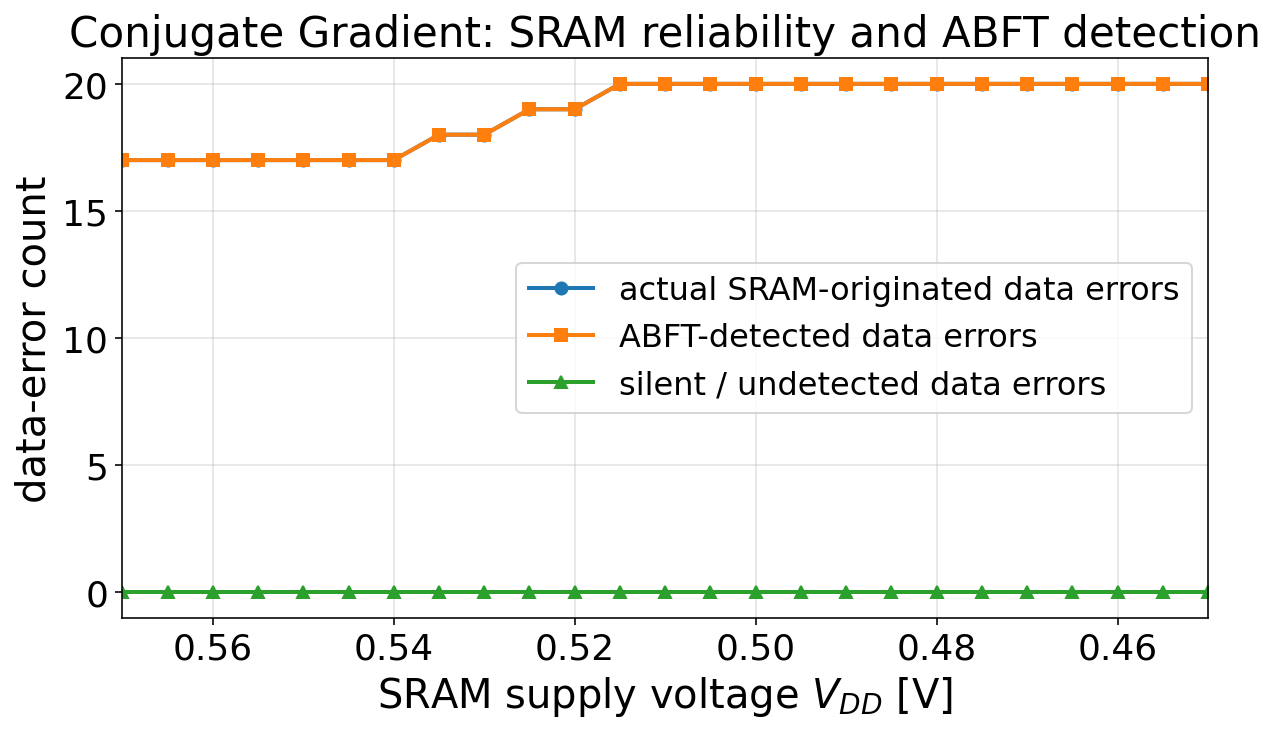}
        \caption{Conjugate Gradient}
        \label{fig:cg_sram_detection}
    \end{subfigure}

    \caption{Final-output checksum-based ABFT detection under
    SRAM-originated faults for (a) GS, (b) ADMIN, and (c) CG.
    Arithmetic fault injection is disabled in this experiment so that
    the effect of SRAM failures can be evaluated independently.}

    \label{fig:sram_detection}
\end{figure*}

The SRAM fault study is evaluated separately from the arithmetic fault
analysis so that the impact of memory-originated errors can be isolated.
In this experiment, arithmetic fault injection is disabled and SRAM
failures are introduced according to the measured voltage-dependent SRAM
reliability profile. Fig.~\ref{fig:sram_detection} shows the number of SRAM-originated final-output
data errors together with the corresponding ABFT-detected and silent errors
for GS, ADMIN, and CG. For GS, the detected-error count overlaps the actual SRAM-originated
error count across the evaluated voltage range, and no silent errors are
observed. A similar behavior is observed for CG, where the SRAM-originated
errors are fully identified by the final checksum verification over the
considered operating points. ADMIN shows a slightly different behavior. The actual SRAM-originated error count exceeds the number of errors detected by ABFT. As a result, a small number of errors remain undetected. This indicates that some SRAM-induced
corruptions can propagate to the final recovered vector while still
satisfying the final checksum relation. Overall, the SRAM results show that the final-output checksum mechanism
can detect memory-originated corruption in addition to arithmetic faults. The detection behavior depends on how an SRAM fault propagates through the corresponding detector. This propagation determines whether the fault can be detected at the final recovered output. The results also show that memory-fault sensitivity is detector dependent. GS and CG provide stronger checksum visibility for the considered SRAM faults, whereas ADMIN exhibits a small number of silent escapes. This difference is caused by the way corrupted stored values influence later detector updates. Therefore, SRAM reliability should be evaluated together with the algorithm-specific error-propagation behavior.

\subsection{Energy and ABFT Overhead}
\label{subsec:energy_overhead}

\begin{figure}[t]
    \centering
    \includegraphics[width=0.95\columnwidth]{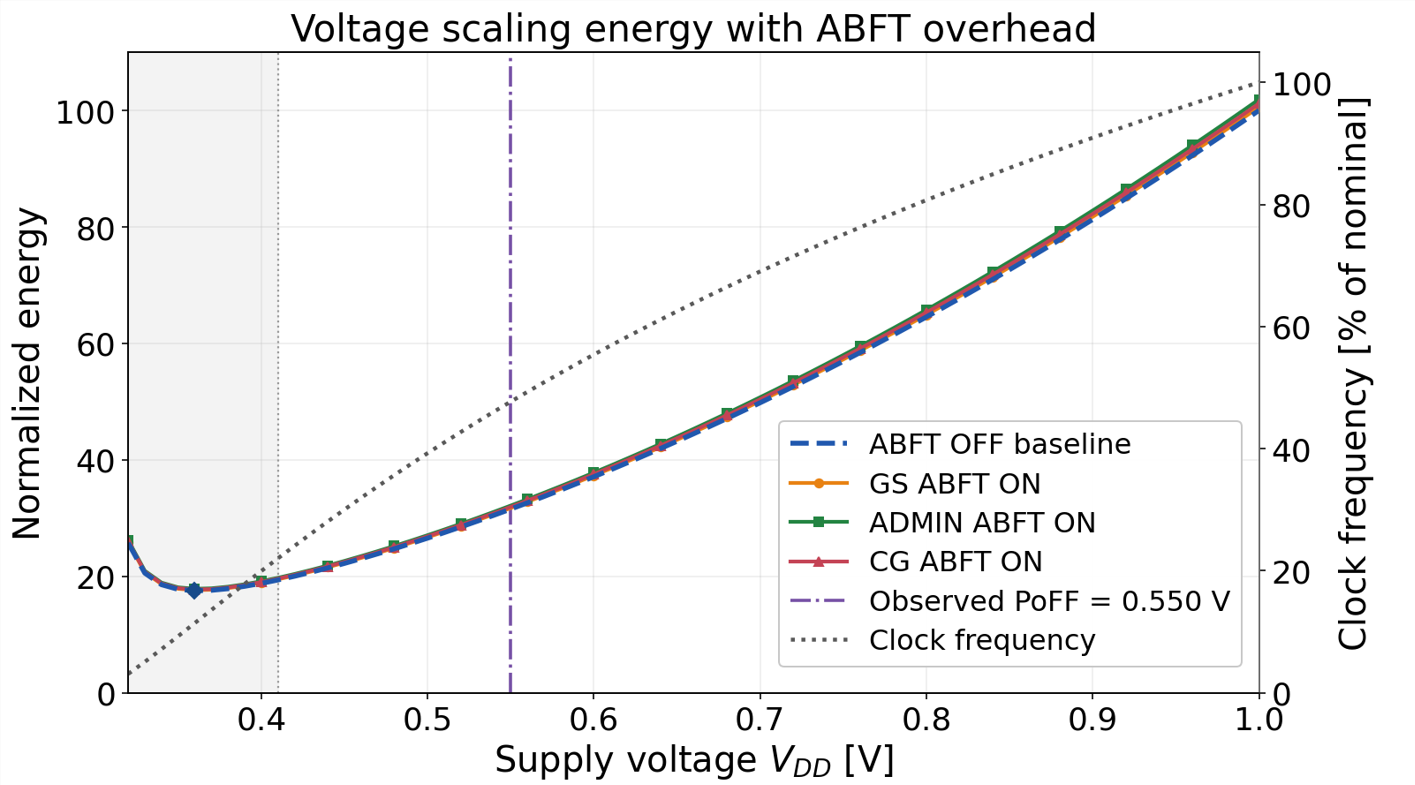}
    \caption{Normalized estimated energy versus supply voltage with and
    without checksum-based ABFT overhead. The reported values are analytical
    workload-based estimates rather than direct hardware energy measurements.}
    \label{fig:energy_vdd}
\end{figure}

\begin{figure}[t]
    \centering
    \hspace*{-0.8cm}\includegraphics[width=0.85\columnwidth]{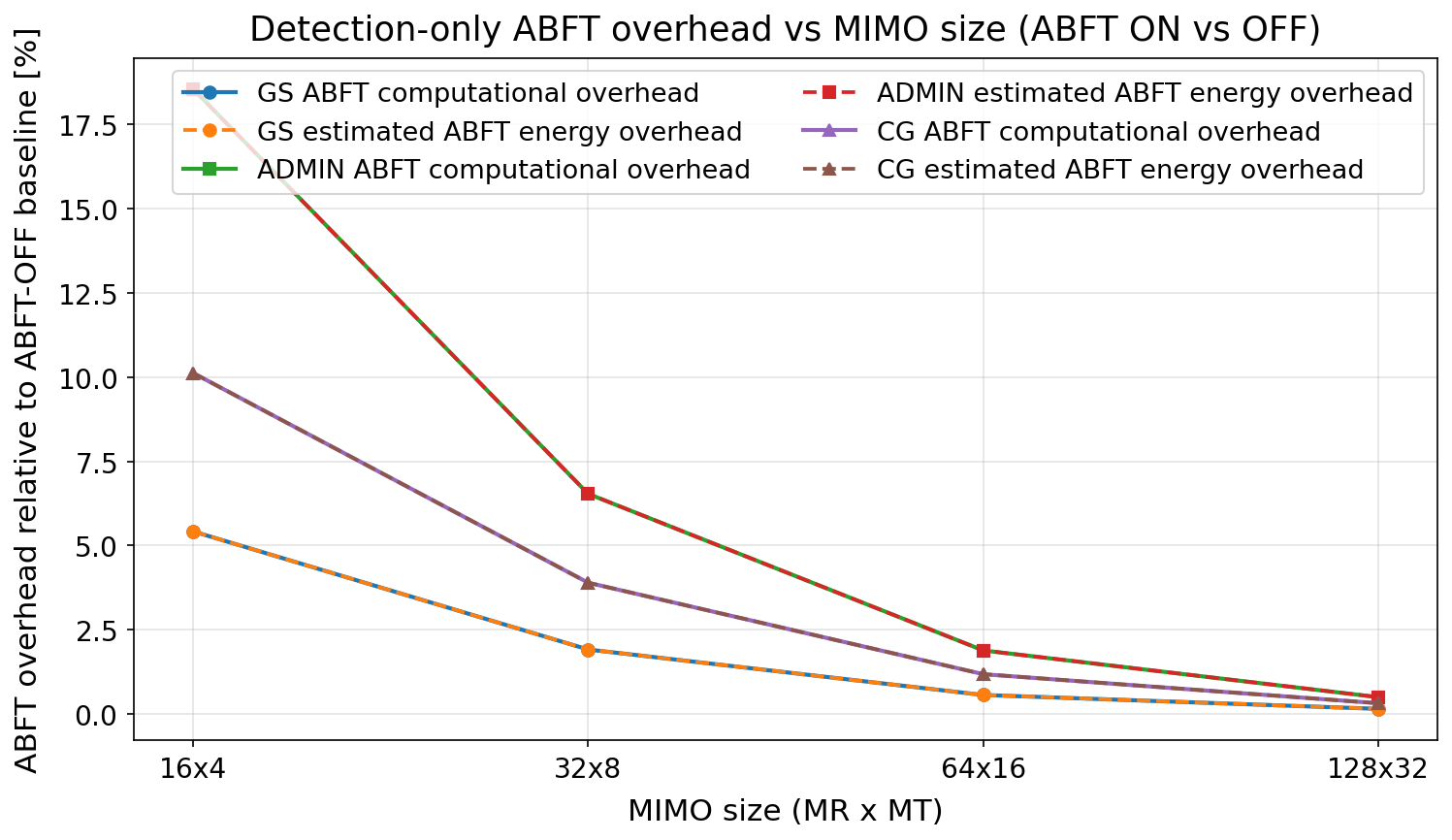}
    \caption{Analytical checksum-based ABFT overhead as a function of MIMO
    system size. The relative overhead decreases as the detector workload
    increases.}
    \label{fig:abft_overhead}
\end{figure}

The energy implications of voltage scaling and checksum-based ABFT are evaluated using analytical workload-based estimates. The reported energy values are therefore not direct hardware measurements. Instead, they are obtained from the detector operation counts together with a voltage- and frequency-scaling energy model. As the supply voltage is reduced, the operating frequency is also reduced to account for the increased circuit delay at lower voltages. The total estimated energy includes both dynamic and static contributions, with the energy values normalized to an ABFT-off operating point at a higher supply voltage and frequency. Since the SPICE-derived reliability models are based on generic Level-1 CMOS devices without a specified fabrication node, the estimated energy savings should be interpreted as normalized, workload-based projections rather than process-calibrated hardware measurements. The absolute energy consumption and the precise voltage-dependent reliability behavior may differ when evaluated using a foundry-specific CMOS technology library. Fig.~\ref{fig:energy_vdd} shows the normalized estimated energy as a function of supply voltage. Starting from the higher-voltage reference point, reducing $V_{DD}$ initially produces a substantial reduction in energy because of the lower dynamic switching cost. However, further voltage reduction requires a lower operating frequency and consequently a longer execution time. In this region, the accumulated static-energy contribution becomes increasingly significant and eventually offsets part of the dynamic-energy reduction. As a result, the estimated total-energy curve reaches a minimum and then increases in the deeper low-voltage region. Thus exhibits the characteristic energy trend expected for near-threshold operation. The ABFT-enabled curves for GS, ADMIN, and CG follow the same overall behavior but remain slightly above the ABFT-off baseline because of the additional checksum-related operations. The relatively small separation between these curves indicates that the ABFT overhead is modest compared with the energy reduction. It is obtained by moving from the higher-voltage operating point toward the near-threshold region. Fig.~\ref{fig:abft_overhead} shows the relative computational and estimated energy overhead introduced by the detection-only ABFT mechanism as the MIMO system size increases. The relative overhead decreases for all three detectors as the system dimension grows because the checksum-related operations represent a progressively smaller fraction of the overall detector workload. Although the exact overhead differs among GS, ADMIN, and CG because of their different arithmetic workloads, the same decreasing trend is observed for all three detectors. Overall, the results indicate that near-threshold voltage and frequency scaling can provide substantial energy savings relative to a higher-voltage operating point, while final-output checksum-based ABFT introduces only a limited additional energy cost.

The proposed scheme supports bounded approximate computing by maintaining the error rate below a predefined limit. As demonstrated for MIMO, brief errors near the point of first failure (PoFF) are tolerable \cite{aliagha2025scissors}.

\section{Conclusion}
\label{sec:conclusion}

This work presented a cross-layer reliability analysis for ABFT-integrated massive MIMO to enable near-threshold MIMO computing. Massive MIMO detection using GS, ADMIN, and CG detectors were considered. Circuit-level PSPICE characterization of the memory and data-path was used to develop voltage-dependent vulnerability profiles. These profiles were then mapped to the corresponding detector operations through workload-aware fault injection. This approach captures both the nonuniform vulnerability of
different bit positions and the different arithmetic exposure of the three
detectors. A detection-only checksum-based ABFT mechanism was used to monitor the
final recovered detector output. The results show that final-output
computational errors begin to appear near the observed arithmetic PoFF of
$V_{DD}=0.550$~V. Although the compute-error rate increases below this operating point, the corresponding BER degradation remains relatively small near the onset of failure. This indicates that corruption of the recovered soft vector does not always result in an incorrect hard decision. The results demonstrate that GS, ADMIN, and CG exhibit different BER sensitivities under identical circuit-derived fault conditions. These differences arise from their distinct arithmetic workloads and error-propagation characteristics. The checksum mechanism detects the majority of final-output arithmetic errors and effectively identifies SRAM-originated corruption. However, a small number of silent errors may occur when the final checksum relation remains satisfied. Finally, analytical workload-based energy estimates show that voltage
scaling provides a substantial reduction in normalized dynamic energy. Meanwhile, the additional energy overhead introduced by checksum-based ABFT remains relatively small. The relative ABFT overhead also decreases as the MIMO
system dimension increases. Overall, the results demonstrate the combined effectiveness of circuit-derived reliability characterization, workload-aware fault injection, and final-output checksum verification. These techniques enable the evaluation and improvement of reliability in energy-efficient near-threshold massive MIMO computing.

\bibliographystyle{IEEEtran}
\bibliography{references}

\end{document}